\documentclass[aps,prd,reprint,nofootinbib]{revtex4-2}
\DeclareUnicodeCharacter{2009}{\,}
\usepackage{xspace}
\usepackage{amsmath}
\usepackage{graphicx}
\usepackage[justification=justified, singlelinecheck=off]{caption}
\usepackage{csvsimple}
\usepackage{booktabs}
\usepackage{nicematrix}
\usepackage{orcidlink}
\usepackage{subcaption}
\graphicspath{{./figures/}}
\usepackage{ragged2e}
\usepackage[normalem]{ulem}
\usepackage[utf8]{inputenc}
\usepackage{todonotes}

\begin{document}
\preprint{}

\title{Is the Binary Black Hole Population Inference from  Gravitational-Wave Data Robust?}
\author{Upasana Das \orcidlink{0009-0009-0307-2265}}
\email{upasana.das@ligo.org} \email{upasana806@gmail.com}
\affiliation{%
School of Physical Sciences, National Institute of Science Education and Research, Bhubaneswar 752050, India
}%

\author{Suvodip Mukherjee \orcidlink{0000-0002-3373-5236}}%
 \email{suvodip@tifr.res.in }
\affiliation{%
Department of Astronomy and Astrophysics, Tata Institute of
Fundamental Research, 1, Homi Bhabha Road, Mumbai-400005,
Maharashtra, India
}%



\begin{abstract}
Gravitational-wave observations are playing an instrumental role in understanding the population of binary compact objects in the Universe. These observations have begun to hint at the mass distribution of binary black holes (BBHs), with tentative evidence for features in the mass distribution beyond a simple power-law. Such features, hence, can be connected with different formation scenarios of BBHs and lead to important astrophysical conclusions. However, it is crucial to understand whether these features are truly astrophysical or connected with any unknown systematics. We show in this work that waveform modelling uncertainties can significantly distort inferred features in the BBH mass distribution, which can be more pronounced than the statistical uncertainty, even with the current generation detectors, which can peak close to the lower edge of the pair instability supernovae (PISN) mass gap, and also can impact the slope of the power-law distribution. So, in order to have a confirmed detection of astrophysical features in the BBH mass distribution and connecting them with BBH formation channels, it is important to consider waveform systematics in the astrophysical population analysis. We show the typical scaling of the systematic error and discuss a few avenues to mitigate this effect for robust measurements in the future. 

\end{abstract}

\maketitle
\section{Introduction}%
\label{sec:Introduction}

It has been a decade since the landmark first detection of gravitational waves (GWs), opening doors to a new way of observing the universe and beginning a new era of multi-messenger astronomy using GWs \cite{Abbott1, Abbott2, Abbott3, Abbott4, Abac1}. Since this first discovery, numerous coalescing compact binaries (CBCs) have been detected by the LIGO-Virgo-KAGRA (LVK) collaboration \cite{GWdetNetwork, LIGO, Aasi_2015, Virgo, Virgo2, Kagra, kagra2, Kagra3}, with the catalog poised to explode further with upcoming observatories such as the LIGO-India \cite{LIGOIndia, LIGOIndia2}, Einstein Telescope \cite{ET, ET2}, Cosmic Explorer\cite{CE, CE2}, and LISA \cite{LISA}. The LVK collaboration \cite{LVK1, LVK2} has catalogued about 218 GW events from CBCs, with the vast majority originating from binary black hole (BBH) systems. This dramatic expansion from detecting a handful of events per year to hundreds has provided original insights into the strong-field nature of gravity, the astrophysical population of compact objects, and the expansion history of the universe\cite{Hubble, GR1, GR2}. The scientific potential of these observations relies critically on our ability to extract physical properties from the GW data, which requires matching the observed strain data against theoretical waveform templates \cite{MatchedFilter}. Any inaccuracy in the modelling of the GW signal can therefore cause incorrect inference of the physical system. 

The modelling of the GW signal has been extraordinarily successful in capturing the emission of GW from coalescing binaries with the aid of multiple techniques ranging from analytical calculations to numerical relativity \cite{PE_intro, PE_intro2, PE_intro3}. The accuracy of these different techniques has played an instrumental role in discovering and characterizing CBC sources. However, in order to achieve accurate astrophysical measurements, continued improvement in the waveform modelling is one of the key requirements \cite{wfAccuracy1, wfAccuracy2}. The fidelity of the waveform model depends on several factors, including GW source properties such as component masses, spin, eccentricity, and orbital precession, as well as the stage of the coalescence (inspiral, merger, and ringdown) that dominates the signal within the detector's sensitive band.

Although the inspiral domain of the signal can usually be modelled at the desired accuracy for the current detectors using approximate waveform techniques such as IMRPhenom and effective-one-body \cite{IMRPhenom1, IMRPhenom2, SEOB1}, the merger and ringdown phase is usually difficult to capture with high accuracy using approximate techniques, and one needs to rely on numerical relativity waveforms, which are accurate but computationally costly. As a result, the parameter estimation of the GW source properties from the data in a Bayesian framework relies on approximate waveforms, as an optimization between accuracy and computing cost. Though for the sources for which the in-band GW signal is dominant by inspiral stage of the signal (primarily low-mass sources), the approximate waveforms perform very well in comparison to the numerical relativity waveforms \cite{NRSur} for most of the parameter space, for the scenarios when the binary spends short time in-band, with a dominant contribution coming from merger and ringdown phase (primarily for high mass events), the waveform inaccuracy can be a vital issue over statistical uncertainties \cite{stat1, stat2}. As a result, understanding the impact of the source-property dependent waveform inaccuracy in the inference of the astrophysical population of binary compact objects is crucial.

One of the main scientific outcomes from the growing BBH catalogue is the reconstruction of the underlying mass distribution $p(m_1, m_2)$ of the component masses $m_1$, and $m_2$, which encodes signatures of the stellar processes that form BBHs  \cite{popn,popn2,popn3,popn4,Afroz:2024fzp} Current analyses reveal structure beyond a simple power law: a low-mass peak near $10M_\odot$ reflecting the mass scale set by core-collapse supernovae, along with possible features near$\sim 20M_\odot$, a prominent peak at $35M_\odot$ widely attributed to pile-up from pulsational pair-instability supernovae (PPISN) \cite{PPISN, PPISN2, PPISN3, Wang:2022gnx, 2021ApJ...913...42W,Karathanasis:2022rtr,Afroz:2025ikg,Antonini:2025ilj, Tong:2025wpz} and a potential high-mass feature at $60M_\odot$ \cite{60Msun,barry}, followed by a suppression above $\sim 80-100M_\odot$ \cite{massGap, massGap2}. However, the reality, sharpness, and location of the two higher mass peaks remain debated due to statistical limitations ($2-3\sigma$ evidence, low event count \cite{ PPISNdoubt2}), astrophysical degeneracies (PPISN vs dynamical formation channels both predicting structure \cite{subpop, ppisndoubt}), and the systematic biases that we aim to address.

These features are inferred through hierarchical Bayesian analyses that combine per-event posterior samples to constrain the parameters of a population model. The accuracy of the individual event posteriors therefore, directly impacts such analyses. Accordingly, the reliability  of gravitational waveform models has been studied extensively at the event level. \citet{Hu} quantified the accuracy of waveform model pairs using their difference, finding that several high-SNR events in GWTC-2.1 and GWTC-3 fail the accuracy criterion. \citet{Purrer2020},  showed how inferred binary parameters for individual events (like GW150914) are impacted by inaccuracies in waveform models. \citet{Owen2023} demonstrated that PN truncation errors, even in the inspiral phase, induce systematic errors in the phase of GWs.  \citet{eventLevel1} compared the \texttt{IMRPhenomXAS} and \texttt{IMRPhenomD} waveform models and reported that for high SNR ($ > 100$) sources, $\sim 3 \%$ to $20\%$ of events will suffer statistically significant paramter biases and that to keep biases $\leq 1\sigma$ for $99\%$ of these loud detections, the average mismatch between waveform models needs to be $\mathcal{O}(10^{-5})$, which is an order of magnitude stricter than current waveform accuracies. Another work by \citet{eventLevel2} treated \texttt{SEOBNRv5PHM} and recovered using \texttt{IMRPhenomXPHM} to find that systematic biases strongly increase with total mass, binary asymmetry, and spin-precession (see also \cite{moreEvent1, moreEvent2, MoreEvent3} for further studies of waveform systematics at the event level.   

Despite this substantial work at the event-level, the propagation of waveform systematics to population-level inference remains a subject of future study to date. Although some studies, such as Ref. \cite{Purrer2020}, do briefly demonstrate that individual measurement errors sum up to a sizable population bias, a thorough quantification has not been done. This gap is concerning because population properties are being increasingly used for GW science, like constraining stellar evolution models, probing formation channels, and measuring cosmological parameters. More recently, for one of the LVK events GW231123 \cite{2025ApJ...993L..25A, 231123systematics}, the variation of the inferred source parameters with waveform was prominent. 

 So, the key question motivating this work is whether the observed features in the BBH mass distribution are influenced, or even driven, by waveform systematics. These biases are more significant for higher-mass BBH systems because their signals are shorter and hence have barely any inspiral, which is better approximated and thus modelled. Though in this work, our primary focus is on the mass distribution of BBHs and how robust it is against waveform systematics, it is also important to study this for other source parameters, such as spin, and how waveform systematics can influence correlated errors in both masses and spin parameters. We defer this study to future work. 

This work aims to quantify the systematic errors arising from the inaccuracies in waveform modelling on the BBH mass distribution, especially when analyzing high-mass BBH systems. The rest of the paper is structured as follows: Section \ref{sec:Waveform Models and Waveform Systematics} details the theory behind GW signals, along with their morphological dependence on the binary's mass, and introduces the waveform families employed in this study. Section \ref{sec:3} represents our core methods and results, moving from individual-event injection-recovery tests discussed in \ref{sec:3A} to hierarchical population inference on realistic mock catalogs in \ref{sec:3b}. In Section \ref{sec:Discussion}, we discuss the broader implications of the systematic biases we found and how they can be accounted for in future observing runs. Finally, Section \ref{sec:Conclusion} summarises our findings.

\section{Inference of astrophysical properties from GW data}%
\label{sec:Waveform Models and Waveform Systematics}

Extracting the physical properties of a coalescing BBH merger from noisy detector data relies fundamentally on Bayesian statistics, where the observed strain data is cross-correlated against a large set of theoretical waveform templates. Consequently, the accuracy of the source parameters depends directly on the fidelity of the waveform models used. In this section, we provide a brief overview of the GW signal morphology. We have discussed different waveform models used in this analysis in Appendix \ref{app:waveform}. 

\subsection{Gravitational-wave signal}
The GW signal from a BBH coalescence is characterised by a set of 15 parameters, comprising 8 intrinsic parameters that describe the physical nature of the source and the remaining extrinsic parameters that describe the source's relation to the observer. The intrinsic properties are defined by the component masses $m_i$ (i = [1,2]) (where we adopt the convention $m_1 \ge m_2$), and the dimensionless spin vectors $\chi_i = \mathbf{S}_i / m_i^2$. From these, the total mass, mass ratio, and chirp mass are defined as \cite{PE1}
\begin{equation}
    M = m_1 + m_2, \quad q = \frac{m_2}{m_1} \leq 1, \quad 
    \mathcal{M} = \frac{(m_1 m_2)^{3/5}}{(m_1 + m_2)^{1/5}}.  
\end{equation}
\begin{figure}[htbp]
    \centering
    \includegraphics[width=\linewidth]{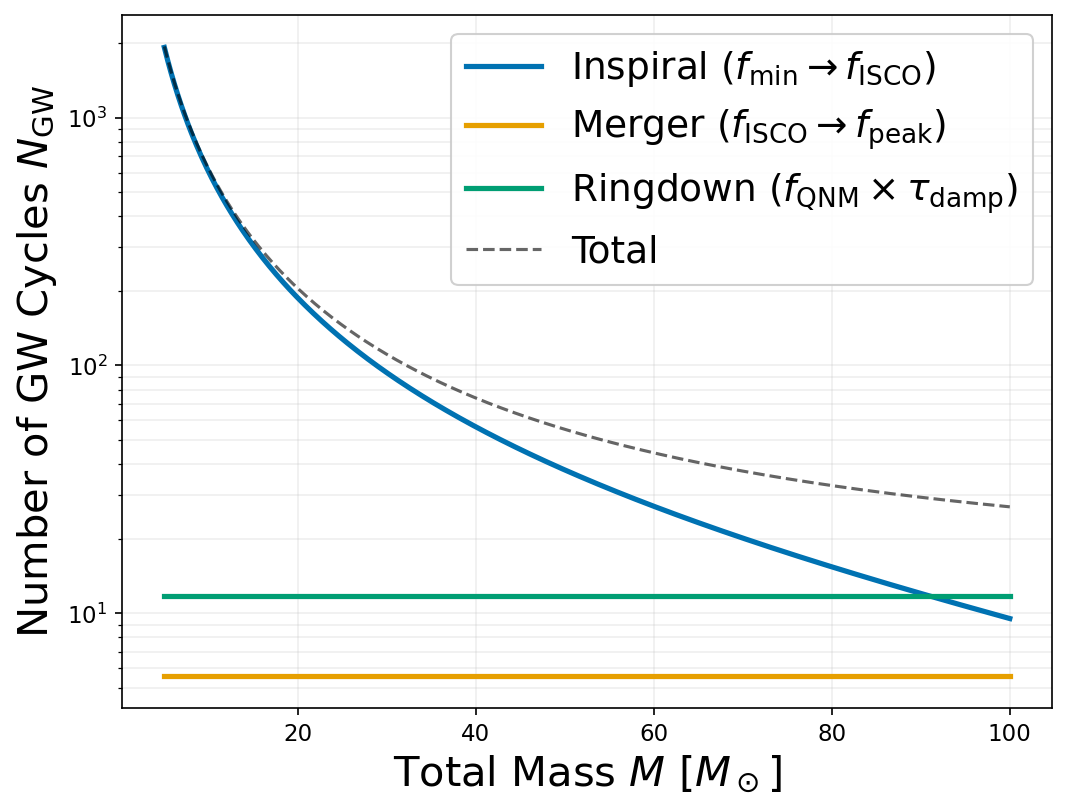}   \caption{\justifying{\textbf{Number of GW cycles per 
    coalescence phase as a function of total mass}, for 
    equal-mass, non-spinning BBH systems ($q = 1$, $\chi = 0$) observed above $f_{\min} = 20\,\mathrm{Hz}$. The inspiral (blue) is computed using post-Newtonian expressions up to the innermost stable circular orbit (ISCO), the merger (red) spans from ISCO to the peak GW amplitude, and the ringdown (green) is estimated from the dominant quasi-normal mode 
    damping time. The total cycle count is shown as the grey dashed curve. For $M \gtrsim 60\,M_\odot$, the inspiral contributes fewer cycles than the ringdown, and the signal is dominated by the merger and ringdown phases, where the waveform 
    models are least accurate.}}
    \label{fig:gw_cycles}
\end{figure}
The extrinsic parameters include the luminosity distance $D_L$, the sky coordinates (right ascension $\alpha$ and declination $\delta$), the inclination angle $\iota$ between the total angular momentum and the line of sight, the polarisation angle $\psi$, the coalescence time $t_c$, and the orbital phase at coalescence $\phi_c$. The parameters $D_L$
and $z$ are connected by the cosmological model, which we take to be the Planck18 results \cite{Planck18}. The recorded strain $h(t)$ in a detector is a linear combination of the two independent GW polarisations, $h_+$ and $h_\times$:
\begin{equation}
    h(t) = F_+(\alpha, \delta, \psi) h_+(t) + F_\times(\alpha, \delta, \psi) h_\times(t),
\end{equation}
where $F_+$ and $F_\times$ are the antenna response functions representing the detector's sensitivity to the respective polarisations, which depend upon the sky location (RA $\alpha$ and Dec $\delta$) and the polarisation angle of the wave ($\psi$). The detector-frame  and the source-frame masses are related as $m_{i, \text{det}} = m_{i, \text{source}} (1+z)$, where $z$ is the redshift of the source. Throughout this work, we quote detector-frame masses unless stated otherwise.

Amongst all source parameters, the total mass $M$ decides which phases (inspiral, merger, and ringdown) of the coalescence fall within the detector's sensitive band. As introduced in Sec. \ref{sec:Introduction}, the GW signal is conventionally divided into three phases: the inspiral, during which the BBHs orbit shrinks, losing energy and angular momentum producing the characteristic "chirp" pattern and is well-described by post-Newtonian (PN) theory; the merger, a highly dynamical strong-field regime that requires NR to solve; and the ringdown, in which the remnant black hole relaxes to a Kerr state and emits in the form of damped quasinormal modes. The frequency at which the system transitions from inspiral to merger scales inversely with $M$ as $f_{\mathrm{merger}} \propto M^{-1}$, 
so that for high mass systems, the inspiral phase may lie partially or entirely outside the detector's sensitive band accessible from LVK. The observable signal is then dominated by the merger and ringdown, making such systems more susceptible to waveform systematics. 

The impact of frequency scaling can be understood directly from the number of observable GW cycles contributed by each phase of the event. Fig.  \ref{fig:gw_cycles} shows the number of GW cycles in the inspiral, merger, and ringdown as a function of M, for equal mass, non-spinning binaries observed from a minimum frequency of 20Hz. The inspiral phase declines steeply with mass, scaling approximately as $N_\text{inspiral} \propto \mathcal{M}^{-5/3}$ from post-Newtonian theory. For a low-mass system, the inspiral contributes more than 1000 observable cycles above $f_{min}$, providing a long information-rich signal. In contrast,  the number of merger cycles is roughly independent of $M$ and is primarily determined by $q$ and spins, and typically lasts for only 0.5 to 1.5 cycles. The number of ringdown cycles is determined by the quality factor of the black hole, which also depends on its final spin.

\begin{figure}[htbp]
    \centering
    \includegraphics[width=0.9\columnwidth]{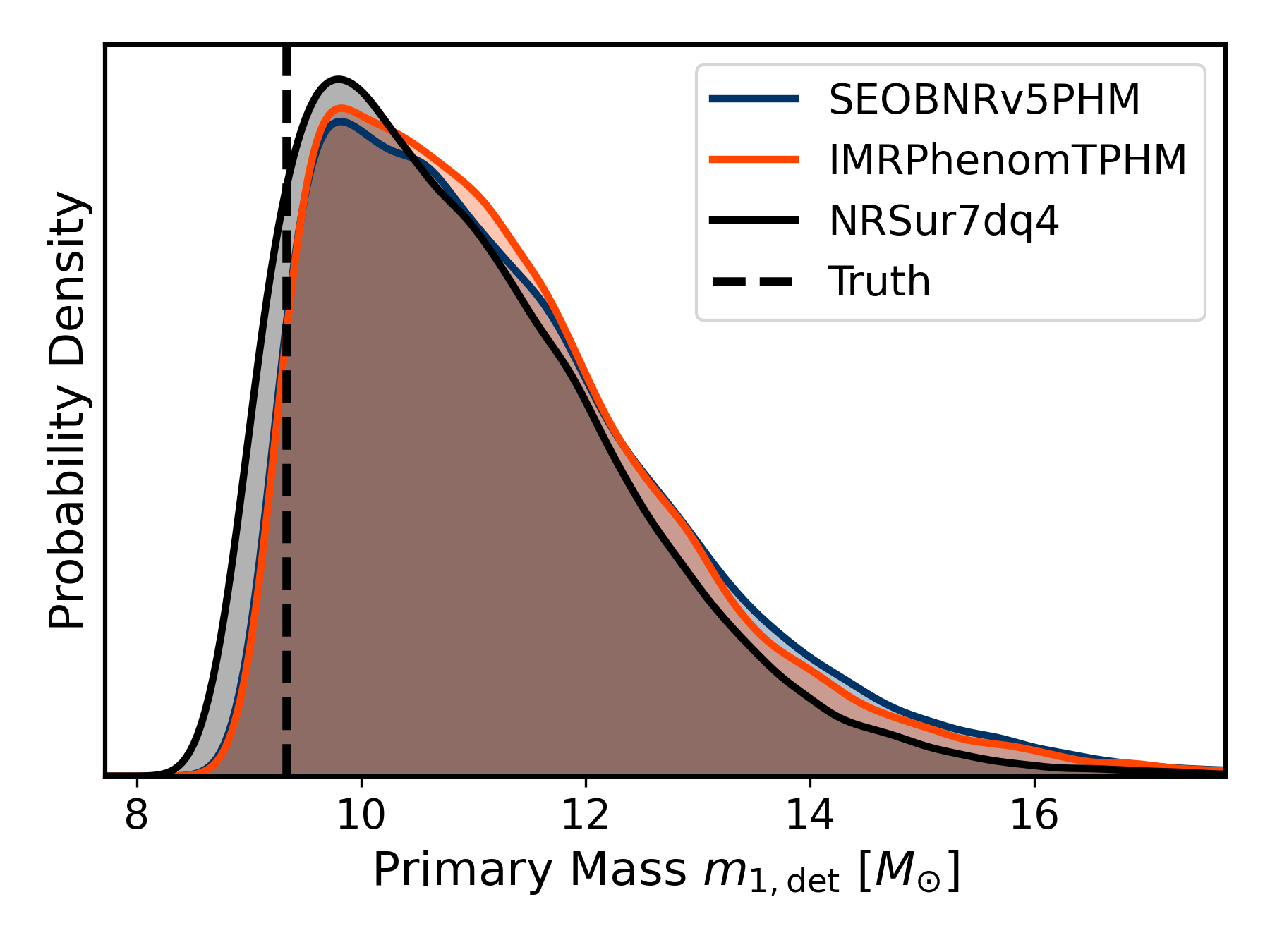}
    
    \includegraphics[width=0.9\columnwidth]{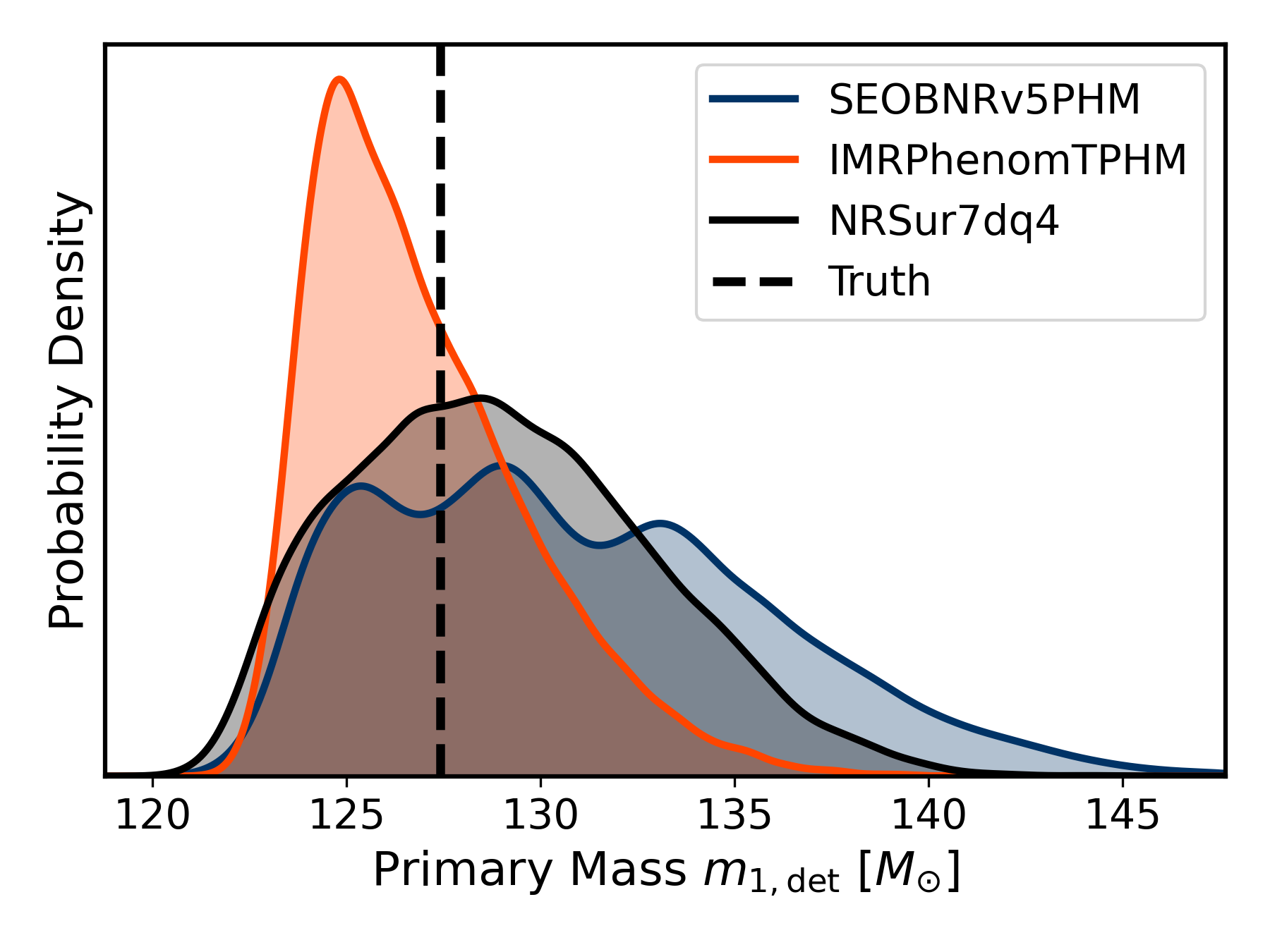}
    
    \caption{\justifying\textbf{Mass-dependent manifestation of waveform systematics.} 1D marginalized posteriors for the detector-frame mass primary($m_{1,\text{det}}$) recovered using \texttt{SEOBNRv5PHM} (blue), \texttt{IMRPhenomTPHM} (red) and \texttt{NRSur7dq4} (black) against an \texttt{NRSur7dq4} injection (black dashed line). \textbf{Top:} For a low-mass system ($M_{\text{tot}} \sim 20 M_\odot$), the inspiral-dominated signal allows for accurate, unbiased recovery. \textbf{Bottom:} For a high-mass system ($M_{\text{tot}} \sim 200 M_\odot$), the merger-ringdown dominance exposes fundamental modelling inaccuracies, resulting in a severe systematic shift away from the true source parameters and a broader posterior. Note that the noise seed is kept fixed for the runs demonstrated in the top and bottom panels.} 
    \label{fig:event_level_motivation}
\end{figure}

The immediate consequences of this can be seen for the recovery of GW source parameters such as component masses. Fig. \ref{fig:event_level_motivation} illustrates the effect with two injection-recovery experiments in which \texttt{NRSur7dq4} serves as the true signal and \texttt{SEOBNRv5PHM}, \texttt{IMRPhenomTPHM}, and \texttt{NRSur7dq4} itself are used for recovery. However, it is important to note that all these waveforms have some level of approximations. Though we consider 
\texttt{NRSur7dq4} as a benchmark waveform model in our analysis, it is only valid for approximately 20 cycles. In our analysis, we change the reference frequency to higher values ($f_{\rm ref} > 20$ Hz) for  total masses less than $60$ M$_\odot$ in our analysis.   
For the low-mass system ($M \sim 20\,M_\odot$), the waveform mismatch is nominal and hence it is possible to recover the source parameters with negligible bias ($<1\sigma)$ and a comparable posterior to the injecting model. For the high-mass system ($M_{\text{tot}}\sim 200\,M_\odot$), where the observable signal is concentrated in the merger-ringdown, the recovered posterior is severely shifted from the injected value with broader posteriors, a direct result of the modelling inaccuracies that dominate these phases. Also, the noise realisation is kept fixed across all runs, and thus, the differences between the low and high mass cases can directly be attributed to waveform systematics rather than statistical fluctuations. This event-level demonstration motivates the central question of this work: how do such mass-dependent biases accumulate across a population of detected events, and can they distort the inferred BBH mass distribution?

\section{Setup to study waveform systematics on GW source population}%
\label{sec:3}

The central aim of this work is to quantify the systematic biases introduced by different waveform approximants and to understand how these biases propagate from the event-level parameter estimation to population-level inference, ultimately impacting astrophysical and cosmological conclusions. Our analysis proceeds in three stages: 
\begin{itemize}
    \item \textit{Controlled injection-recovery tests that isolate waveform systematics as a function of mass,} 
    \item \textit{Construction of a realistic mock BBH catalog incorporating an astrophysical mass distribution and detector selection effects,} 
    \item \textit{Hierarchical population inference to quantify the cumulative distortion of the BBH mass spectrum due to waveform systematics.}
\end{itemize}
In the following sections of the paper,  we describe each of these stages in detail. 


\subsection{Individual mass-bin injection-recovery Framework}
\label{sec:3A}

\subsubsection{Method}
The initial stage of our analysis involves injection-recovery tests to isolate waveform systematics across the mass spectrum. Before assessing the impact of systematics on a full astrophysical population, we need to understand how systematic bias behaves fundamentally at the single-event level. As discussed in section \ref{sec:Waveform Models and Waveform Systematics}, waveform modelling inaccuracies are expected to strongly scale with mass. To map a clear dependence without variables of fluctuating spins, varying mass ratios, or differing detector responses, we start with a fixed parameter injection strategy.


\begin{figure*}[htbp]
    \centering
    \includegraphics[width=\linewidth]{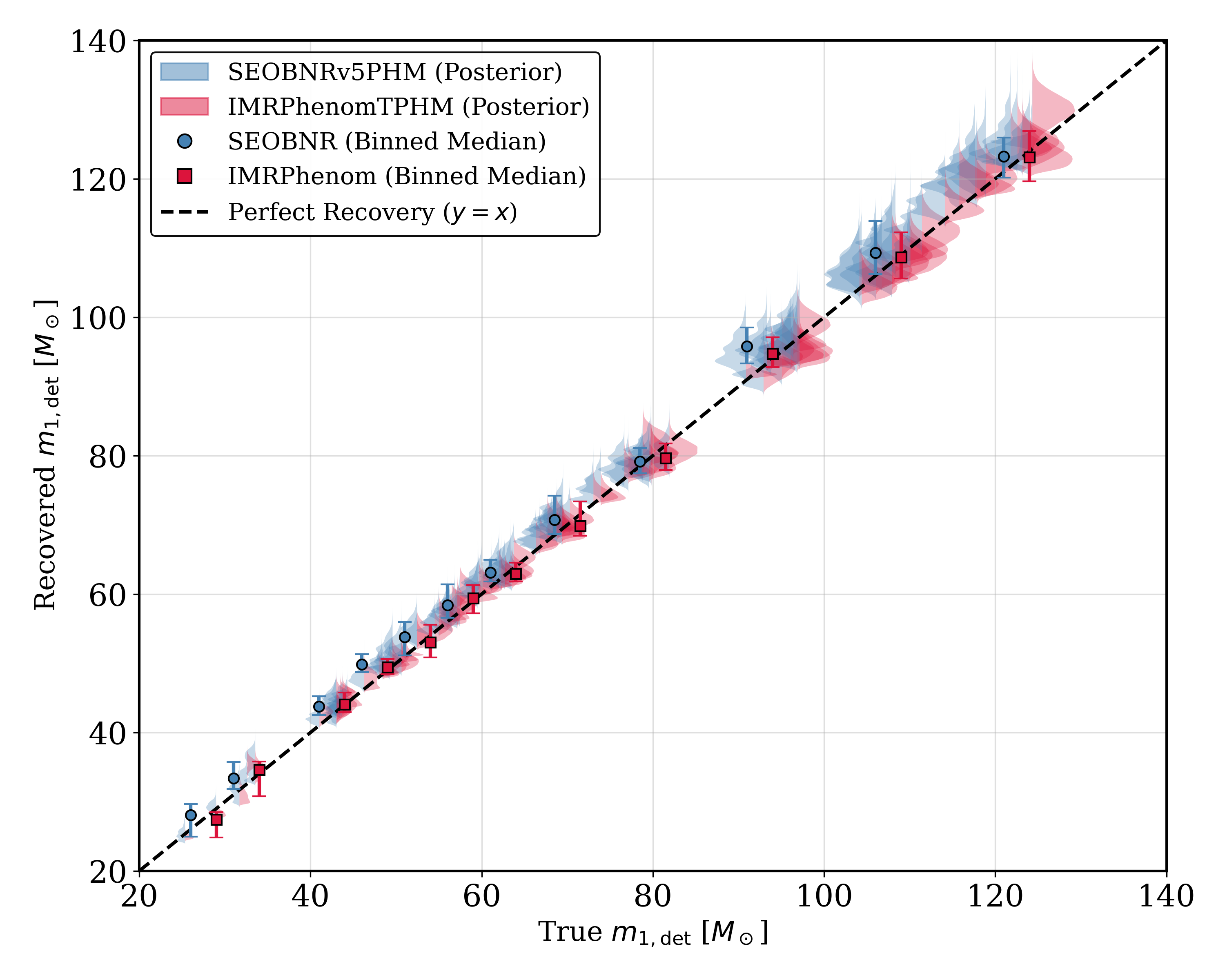} 
    \caption{\justifying Recovered detector-frame $m_1$ plotted against true injected values for signals generated with \texttt{NRSur7dq4} and recovered independently with \texttt{SEOBNRv5PHM} (blue) and \texttt{IMRPhenomTPHM} (red) for our initial fixed bins test case. Violin plots show the full marginalised posterior for each event, while the filled markers (circles for \texttt{SEOBNRv5PHM}, squares for \texttt{IMRPhenomTPHM}) indicate the binned posterior medians. The black dashed line indicates perfect recovery. Error bars span the 68\% credible intervals.}
    \label{fig:rec_vs_true_violin}
\end{figure*}

\begin{enumerate}
    \item We construct a catalog of 110 events to be injected with component masses drawn from fixed mass bins of width ranging between $5-10M_\odot$ between $15-100M_\odot$, which also forced $q \sim 0.8-0.9$. All other intrinsic and extrinsic parameters are held fixed across injections, allowing us to systematically scan the mass range and directly assess the evolution of waveform-induced biases as a function of mass alone, such that changing detector responses also does not affect inference. The lower masses are excluded in this analysis, as the systematic bias for inspiral-dominated events is negligible. 
    
    \item  The true event signals are constructed using the \texttt{NRSur7dq4} surrogate model and injected into a simulated detector corresponding to the H1-L1 detector network. The data in a given detector $I \in\{H1, L1\}$ is denoted by $d^{I}$ can be expressed as  
    \begin{equation}
    d^{I}(t) =  F^{I}_+ h_+(t) + F^{I}_\times h_\times(t) + n^{I}(t),
    \end{equation}
where $F^{I}_{+,\times}$ are the antenna pattern functions and $n^{I}(t)$ represents the noise realizations generated using the predicted O5 power spectral density (PSD) \cite{PSD,psd2,psd3}. To further suppress the contamination from noise fluctuation and hence the impact from statistical fluctuation is limited on the error budget, we have chosen a higher SNR threshold for these studies. This helps in understanding more clearly the effect of systematic errors on the statistical uncertainties. We fix the matched-filter signal-to-noise ratio (SNR) to $\sim 100$ for all injections. This is achieved by adjusting the luminosity distance $D_L$ while holding the sky location ($\alpha, \delta$) and orientation ($\iota, \psi$) fixed.

\item For each event, we recover the source parameters $m_i, q \text{ and } \mathcal{M}$ using bilby \cite{Bilby, Bilby2} with waveform templates from \texttt{IMRPhenomTPHM} and \texttt{SEOBNRv5PHM}. Appropriate priors (check Table \ref{tab:hyperpriors}, 2nd panel) are put in place with phase and time marginalised over. Sampling is carried out using the Nessai sampler. Nessai is the Nested Sampling with AI sampler, which trains a normalising flow to learn likelihood contours during nested sampling and then samples directly from those contours to new samples according to the provided prior \cite{nessai, nessai2, nessai3}. A fixed seed is used throughout the analysis to ensure reproducibility and uniformity.

\item The systematic bias $\Delta$ is defined for each event as the difference between the  median of the recovered posterior distribution and the true injected value:
\begin{equation}
    \Delta_\theta \equiv \text{median}(\theta_{\text{rec}}) - \theta_{\text{true}}. \label{eq:bias_def}
\end{equation}
\end{enumerate}


\subsubsection{Result}

To establish a baseline understanding of waveform modelling systematics at the individual event level, we first examine the direct recovery of the detector-frame primary mass. Figure \ref{fig:rec_vs_true_violin} shows the recovered median $m_{1,\mathrm{det}}$ against the true injected values for a simulated catalog, utilizing our recovery waveforms. 


\begin{figure*}[htbp] 
    \centering
    \begin{minipage}[t]{0.5\textwidth} 
        \centering
        \includegraphics[width=\linewidth]{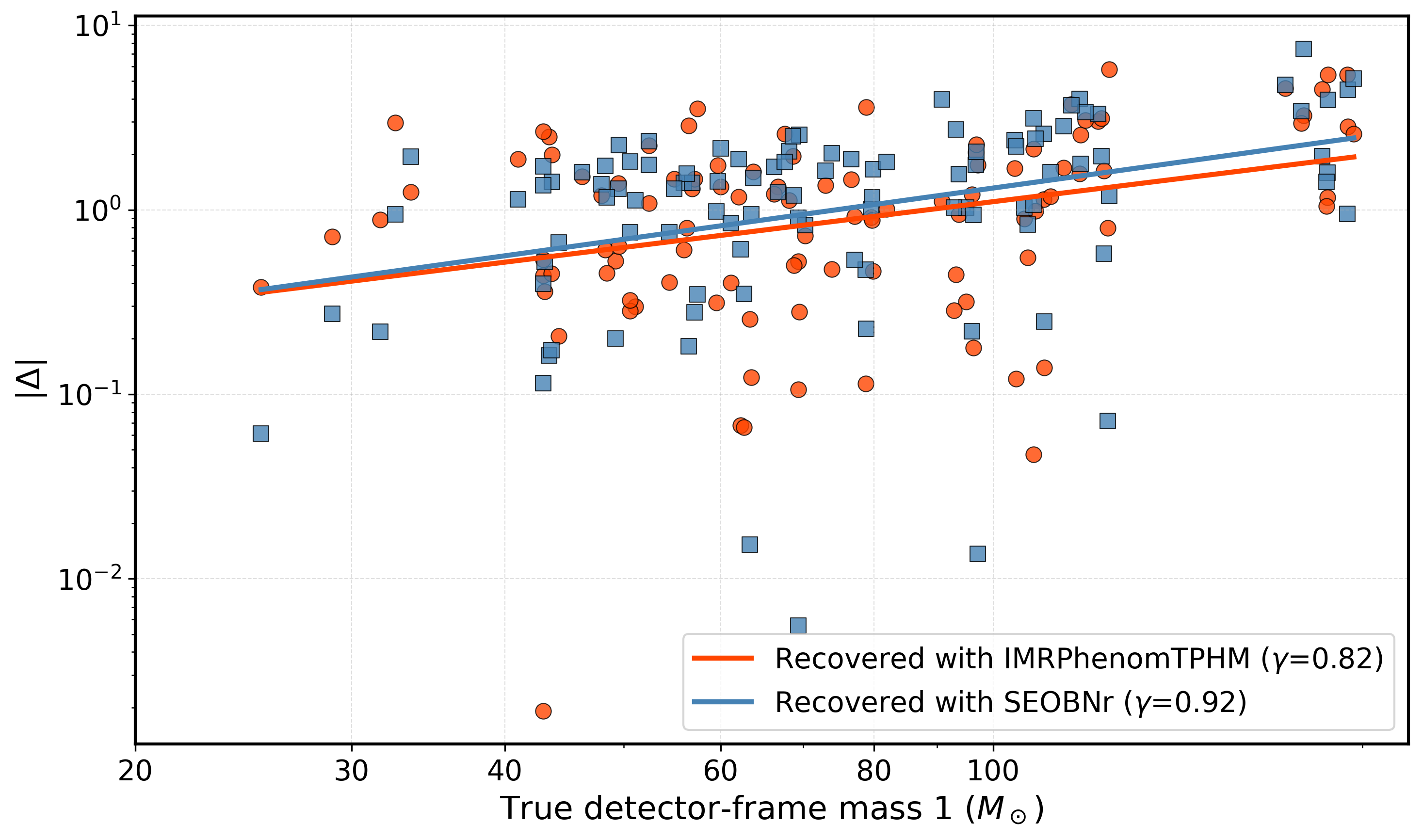} 
        \label{fig:figure1}
    \end{minipage}%
    \hfill 
    \begin{minipage}[t]{0.5\textwidth} 
        \centering
        \includegraphics[width=\linewidth]{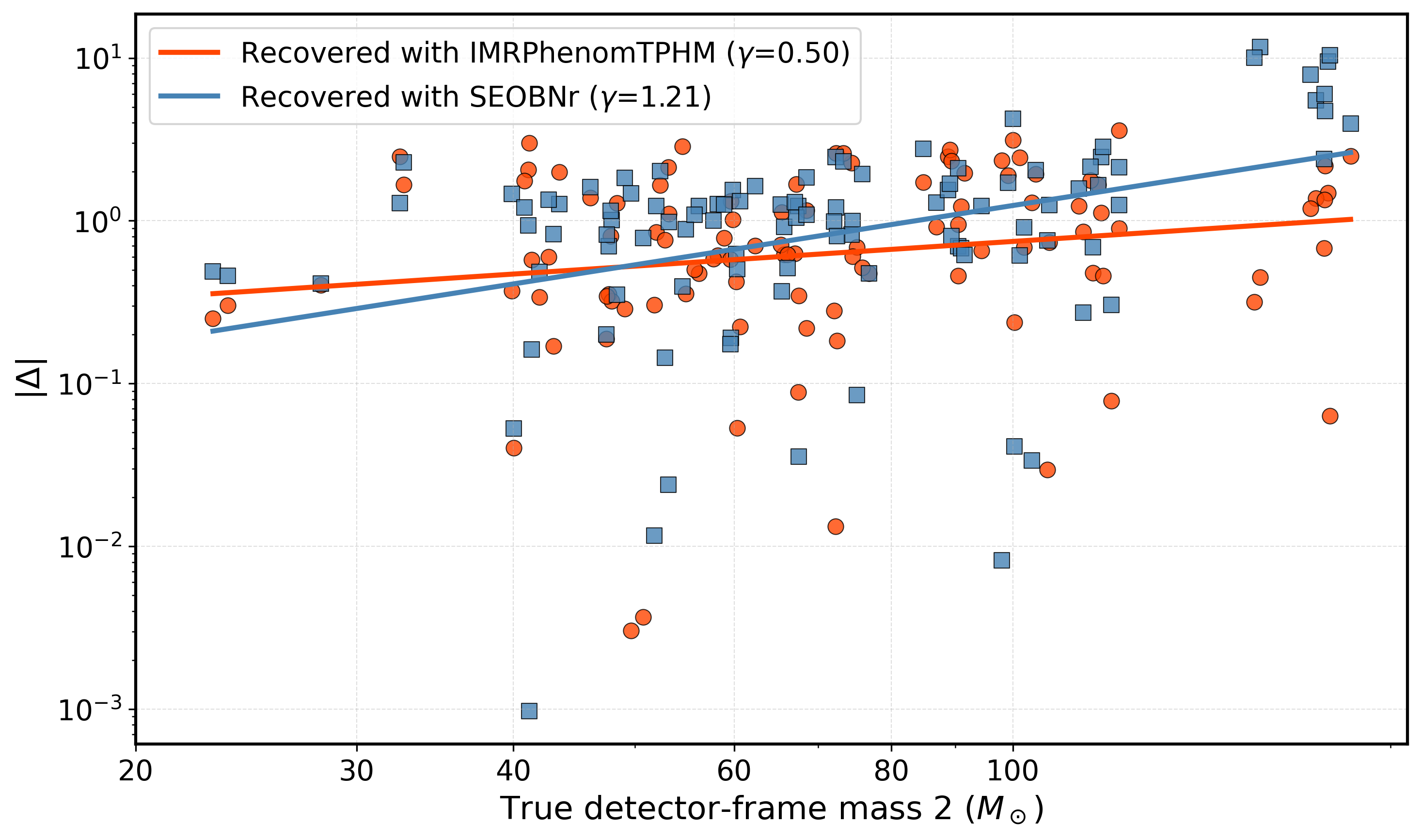}
        \label{fig:figure2}
    \end{minipage}
    \caption{\justifying \textbf{Systematic bias scaling with mass for fixed-bin injections.} The absolute systematic bias $|\Delta|$ between the injected \texttt{NRSur7dq4} signal and the recovered posterior median using using \texttt{IMRPhenomTPHM} and \texttt{SEOBNRv5PHM} is shown as a function of the true mass. The left panel shows this scaling with $m_1$, and the right panel shows the same with $m_2$. The injections were performed in fixed mass bins of width $5M_\odot$ with fixed extrinsic parameters, and the matched-filter SNR is kept constant. The solid lines represent power-law fits ($|\Delta| \propto M^\gamma$) in log-log space, indicating that waveform systematics grow monotonically with the mass of the binary components.} 
    \label{fig:bias_vs_mass_fixedbin}
\end{figure*}

 At lower masses ($m_{1,\mathrm{det}} \lesssim 40 M_\odot$), the recovered medians from both models closely track the line of perfect recovery, and the violin distributions are narrow. As the mass increases, the posteriors progressively broaden, and for massive events ($m_{1,\mathrm{det}} \gtrsim 70 M_\odot$), the bias evolves in a model-dependent way. \texttt{SEOBNRv5PHM} shows a mild positive bias, while \texttt{IMRPhenomTPHM} shows scatter in both directions. This behaviour is also reminiscent of Fig. \ref{fig:event_level_motivation} where the two recovery models, nearly indistinguishable at low masses, diverge at higher masses.


Now, to quantify this trend, Fig. \ref{fig:bias_vs_mass_fixedbin} shows the absolute systematic bias $|\Delta_\theta|$ (Eq.~\ref{eq:bias_def}) as a function of the true injected mass for $m_1 \text{ and } m_2$ obtained from the controlled fixed mass-bin injections described in Section \ref{sec:3A}. As the plots have been made in log-log space, a clear trend consistent, on average, with power-law relation in mass (represented by $M \sim m_i, \mathcal{M}$), is seen, given as  
\begin{equation}
    |\Delta| \propto M^\gamma 
\end{equation}
with best fit slopes obtained from linear regression in log-log space summarised in Table \ref{tab:gamma_slopes}. The slopes are of order unity, indicating an approximately linear scaling of bias with mass.  The physical origin of this scaling was discussed in Sec. \ref{sec:Waveform Models and Waveform Systematics}. 

\begin{table}[htbp]
\centering
\caption{ \justifying Best-fit power-law slopes $\gamma$ for the bias-mass scaling $|\Delta| \propto M^\gamma$, obtained from linear fits in log-log space for the fixed-bin and realistic PL+G catalog.}
\label{tab:gamma_slopes}
\renewcommand{\arraystretch}{1.25}
\begin{tabular}{l c c c c}
\hline\hline
 & \multicolumn{2}{c}{Fixed bins} 
 & \multicolumn{2}{c}{PL+G}  \\
\cmidrule(lr){2-3} \cmidrule(lr){4-5}
Param 
 & \texttt{SEOB} & \texttt{IMRPhenom} 
 & \texttt{SEOB} & \texttt{IMRPhenom}  \\
\hline
$m_1$ & $0.92$ & $0.82$ & $0.81$ & $0.92$  \\
$m_2$ & $1.21$ & $0.50$ & $1.12$ & $1.19$  \\
\hline\hline
\end{tabular}
\end{table}

The slope is steeper for $m_2$ in the fixed-bin setup than for $m_1$ because $m_2$ is intrinsically less well-constrained. While $\mathcal{M}$, the best measured parameter, is determined by the inspiral phase, $q$ (and subsequently $m_2$) derives its constraining power from the relative amplitudes of subdominant harmonics and the merger-ringdown. At high masses, although both measurements are bad, $q$ degrades faster because its primary information content is in the phases where the waveform templates are least accurate. Thus, errors $m_1$ and $q$ compound in the recovery of $m_2$, especially in the high-mass regime.

\begin{figure*}[htbp] 
    \centering
    \begin{minipage}[t]{0.5\textwidth} 
        \centering
        \includegraphics[width=\linewidth]{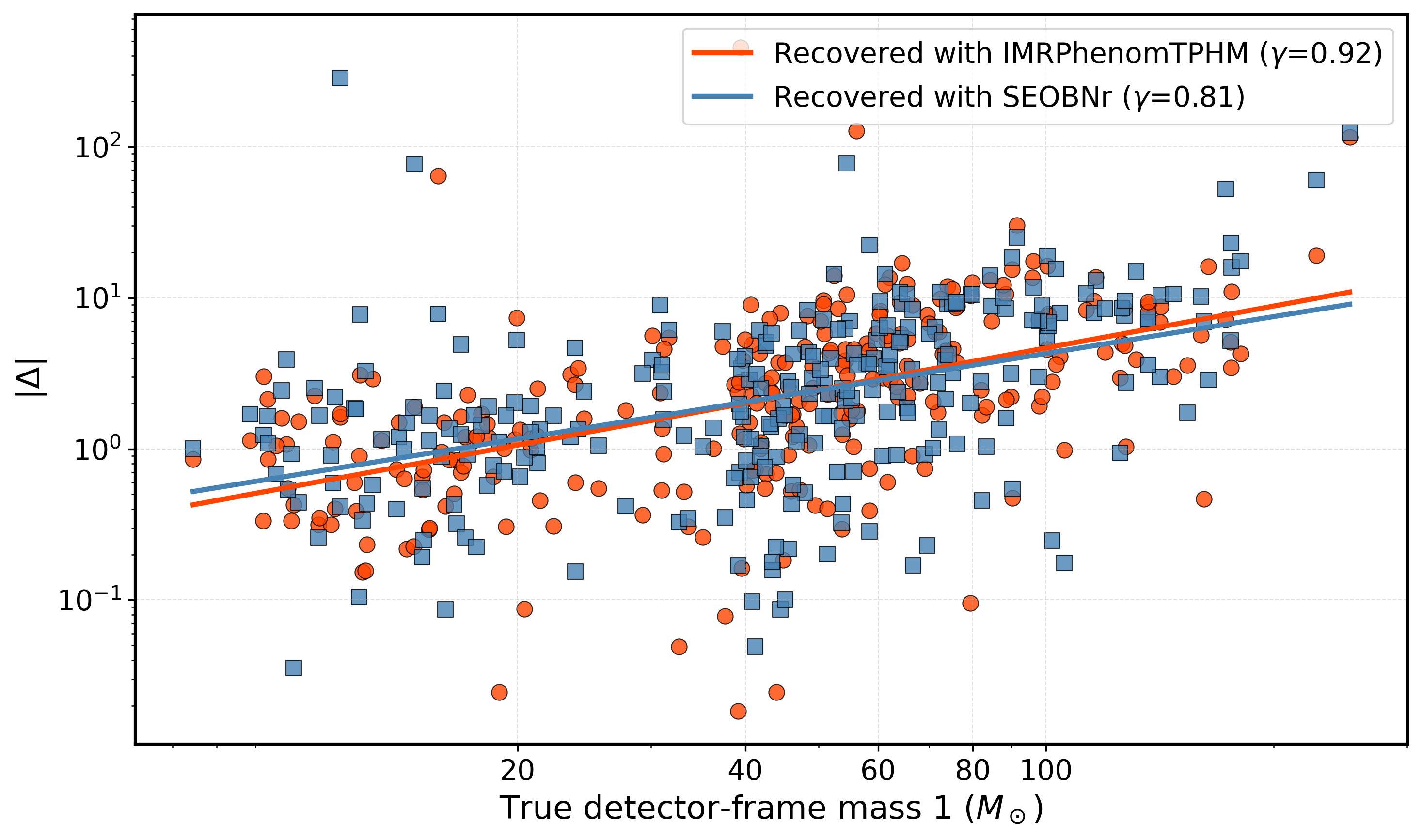} 
    \end{minipage}%
    \hfill 
    \begin{minipage}[t]{0.5\textwidth} 
        \centering
        \includegraphics[width=\linewidth]{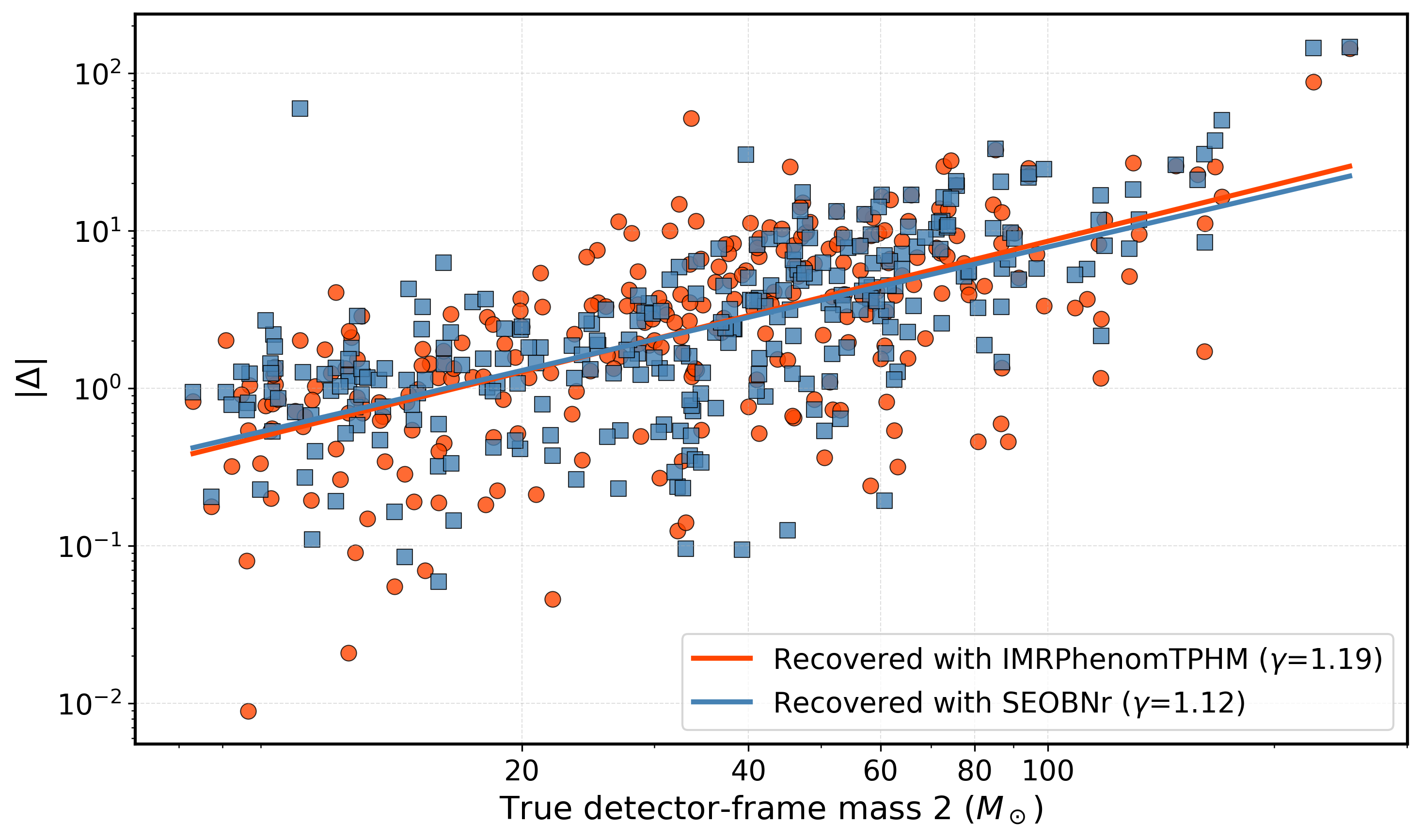}
    \end{minipage}
    
    \vspace{0.5cm} 
    
    
    \caption{Same as Fig. \ref{fig:bias_vs_mass_fixedbin}, but for injection catalog drawn from realistic PL+G (row 1). In this case, all the 15 parameters are allowed to vary according to the methods described in Sec. \ref{sec:3b}}
    \label{fig:bias_scaling_populations}
\end{figure*}

\begin{figure}[htbp]
    \centering
    \includegraphics[width=\linewidth]{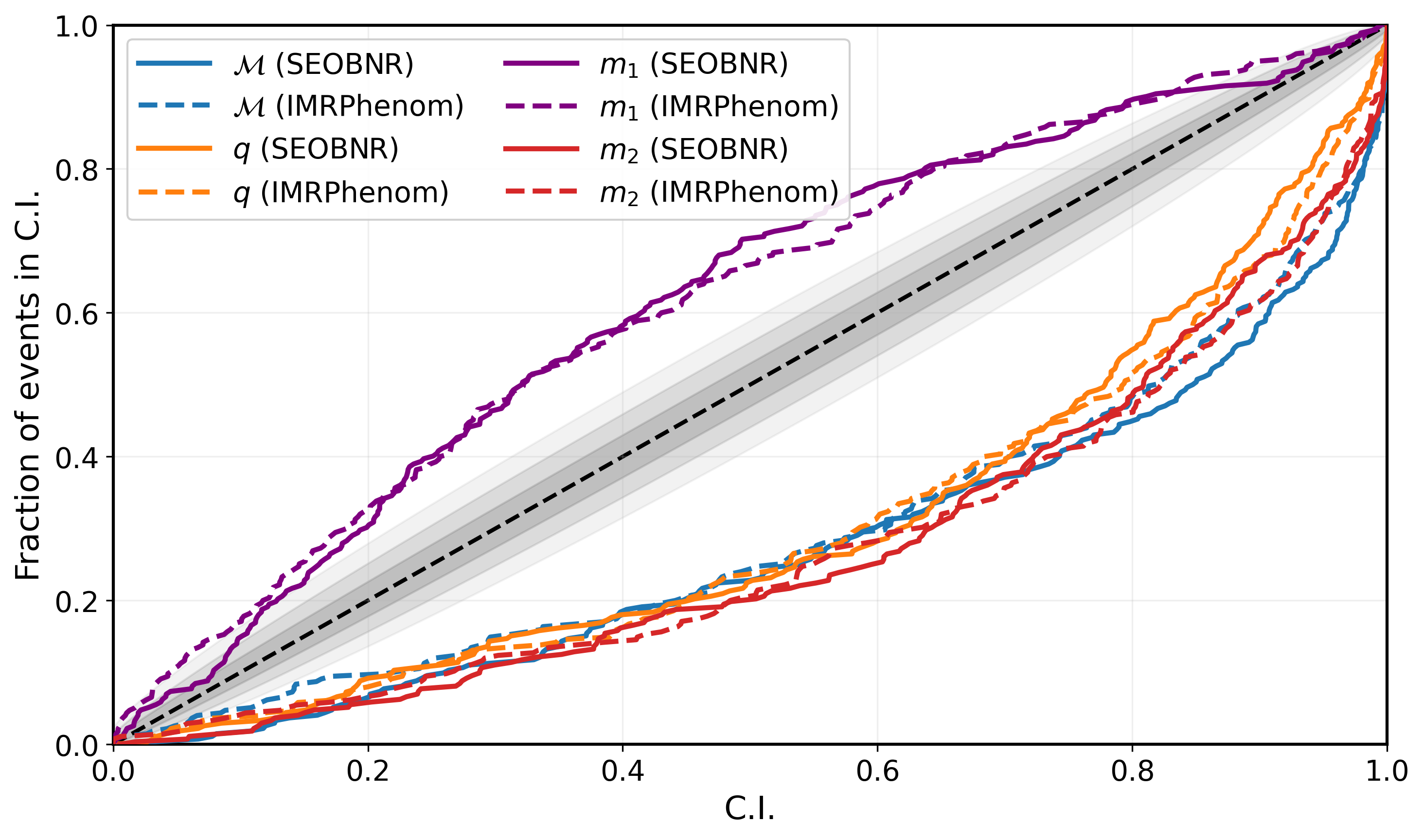} 
    \caption{\justifying \textbf{P-P Plot for recovered source parameters} Each curve shows the fraction of events with credible intervals (C.I.) containing the true value, shown for $\mathcal{M}$ (blue),  $q$ (orange), $m_1$ and $m_2$ for injections drawn from the PLG population and recovered using \texttt{SEOBNRv5PHM} (solid) and \texttt{IMRPhenomTPHM} (dashed).}
    \label{fig:pp_plot}
\end{figure}

\subsection{Generation of Mock Population for an astrophysical population}
\label{sec:3b}
\subsubsection{Method}
Having established the behaviour of waveform systematics at the event level, we extend the study to study the impact on realistic astrophysical 
populations. Using the \texttt{GWSim} \cite{GWSim} package, we simulate a universe consistent with  $\Lambda$CDM cosmology, where BBH merger rates follow the Madau-Dickinson star formation history \cite{MadauDickinson}.  We consider a Power Law + Gaussian (PL+G), a mixture of a truncated power law and a Gaussian component, which is consistent with that used in the LVK population analyses for GWTC-3 \cite{35peak, popn}.  True hyperparameters used to generate the model is listed in Table \ref{tab:true_hyperparams}.


\begin{table}[htbp]
\centering
\caption{True population hyperparameters used to generate the  Power Law+Gaussian
mock catalog.}
\label{tab:true_hyperparams}
\begin{tabular}{lc}
\hline
Parameter & True injected value \\
\hline
$\alpha$     & 3.5 \\
$\beta$      & 1.1  \\
$m_{\min}\,[M_\odot]$  & 5     \\
$m_{\max}\,[M_\odot]$  & 100  \\
$\delta_m\,[M_\odot]$  & 4.8    \\
$\mu_g\,[M_\odot]$     & 34  \\
$\sigma_g\,[M_\odot]$  & 4   \\
$\lambda_g$  & 0.038  \\
\hline
\end{tabular}
\end{table}


Spin magnitudes are drawn from a uniform distribution. The detector selection effects are incorporated using the O5 noise PSD, and an SNR cutoff is set at 12. The resulting detected events constitute the mock catalog used in subsequent analyses. The primary and secondary mass distributions are defined as 
\begin{widetext}
\begin{equation*}
\begin{aligned}
P_1\left(m_1 \mid m_{\min }, m_{\max }, \alpha\right) &= \begin{cases}(1-\lambda_g) \mathcal{P}\left(m_1,-\alpha\right)+\lambda_g G\left(m_1, \mu_g, \sigma_g\right), \quad \text{for } m_{\min }<m_1<m_{\max }, \\
0, \quad \text { otherwise }.\end{cases} \\
P_2\left(m_2 \mid m_{\min }, m_1, \beta\right) &= \begin{cases}\mathcal{P}\left(m_2,-\beta\right), \quad \text{for } \quad m_{\min }<m_2\leq m_1, \\
0, \quad \text { otherwise }.\end{cases}
\end{aligned}
\end{equation*}
\end{widetext}
Here, $\mathcal{P}\left(m_1,-\alpha\right)$ and $\mathcal{P}\left(m_2,-\beta\right)$ denote normalised power-law distributions with spectral indices $\alpha$ and $\beta$ respectively, defined over the specified mass ranges. The term $G\left(m_1, \mu_g, \sigma_g\right)$ represents a Gaussian probability density function with mean $\mu_g \text{ and standard deviation } \sigma_g$. The parameter $\lambda_g \in [0,1]$ represents the Gaussian mixing fraction between the power law and Gaussian components.

Importantly, the bias-mass scaling seen in the previous section persists even in the more general injection settings, where all intrinsic and extrinsic parameters are allowed to vary, and the events are drawn from realistic astrophysical populations. Figure \ref{fig:bias_scaling_populations} shows the results for injections drawn from the PL+G mass model. Despite the additional scatter introduced by varying spins, inclinations, sky positions, and distances, the underlying power-law trend remains clearly visible in both cases, with best-fit slopes consistent with those obtained from the fixed-bin analysis within uncertainties. 

The slopes are shallower than the previous control setup and more similar between $m_1$ and $m_2$. This convergence reflects the additional parameter degeneracies that arise when spins and other extrinsic parameters are free, most notably the spin-mass degeneracy and inclination-distance correlation, which can either amplify or partially compensate the bias for individual events, but the systematic trend persists on average. The increased scatter in $\Delta$ is also visible as approximately one to two orders of magnitude compared to the relatively tight scaling in the fixed-bin case due to the same reason.

As a complementary check to quantify the performance of our parameter estimation \cite{pp_plot}, we verify that the event-level biases are statistically significant across our mock catalogs. Fig. \ref{fig:pp_plot} shows the probability-probability (P-P) plot for the recovered source parameters $m_1, m_2, \mathcal{M}$ and $q$ constructed from the PLG catalog recovered with our chosen recovery models. A "sag" is seen in all the curves, which means that on average, the true parameters are further away from the peak of the posterior than expected, and thus fewer or more injections would be recovered at a given interval level, indicating a bias in the distribution. Specifically, the $m_1$ curve bending upwards means that the injected values consistently lie in the lower percentiles, that is, the lower end of the posterior tail, proving that recovery by both approximate models systematically overestimates $m_1$ for most events. 

\subsubsection{ Population-Level Analysis}
To quantify the cumulative impact of waveform systematics on the inferred BBH mass distribution, we perform Hierarchical Bayesian Estimation \cite{bayesianInf} on the mock catalog described in Sec \ref{sec:3b}, using a similar framework as in ICAROGW \cite{icarogw}. By comparing the recovered population hyperparameters 
$\Lambda_{\text{rec}}$ against the known true injected values $\Lambda_{\text{true}}$, we measure the distortion of population features. 

Let us observe GW data ${x_i}$ for N$_{\text{obs}}$ detected BBH mergers which have source parameters $\theta_i$ drawn from a population model controlled by population hyperparameters $\Lambda$. The hierarchical posterior on the hyperparameters is given by \cite{HBI, HBI2} 

\begin{equation}
p\left(\Lambda \mid\{x\}, N_{\mathrm{obs}}\right) \propto \pi(\Lambda) \prod_{i=1}^{N_{\mathrm{obs}}} \frac{\int p\left(x_i \mid \Lambda, \theta\right) p_{\mathrm{pop}}(\theta \mid \Lambda) d \theta}{\int p_{\mathrm{det}}(\theta, \Lambda) p_{\mathrm{pop}}(\theta \mid \Lambda) d \theta},
\end{equation}
where $\theta$ is the set of individual-event source parameters of the BBH,  $p\left(x_i \mid \Lambda, \theta\right)$ is the single-event likelihood, $\pi(\lambda)$ is the prior on hyperparameters, $p_{\mathrm{pop}}(\theta \mid \Lambda)$ is the population prior on source paramters given $\Lambda$ (Table~\ref{tab:hyperpriors}) and $P_\text{det}$ corrects for the selection function. Thus, the denominator represents the expected number of detections, defined as
\begin{equation} 
\frac{N_{\exp }(\Lambda)}{N(\Lambda)}=\int p_{\operatorname{det}}(\theta, \Lambda) p_{\mathrm{pop}}(\theta \mid \Lambda) \mathrm{d} \theta.
\end{equation}

\begin{table}[t]
\centering
\caption{Priors on population hyperparameters used in the hierarchical analysis and event-level parameter estimation (PE) parameters.}
\label{tab:hyperpriors}
\begin{tabular}{lc}
\hline
Parameter & Prior \\
\hline

\multicolumn{2}{c}{\textbf{Population Hyperparameters}} \\
\hline
$\alpha_0$ & $\mathcal{U}(1,\,8)$ \\
$\beta_0$ & $\mathcal{U}(-2,\,8)$ \\
$m_{\min}$ & $\mathcal{U}(2,\,10)$ \\
$m_{\max}$ & $\mathcal{U}(60,\,150)$ \\
$\mu_g$ & $\mathcal{U}(20,\,50)$ \\
$\sigma_g$ & $\mathcal{U}(1,\,15)$ \\
$\lambda_g$ & $\mathcal{U}(0,\,1)$ \\

\hline
\multicolumn{2}{c}{\textbf{Event-level PE Parameters }} \\
\hline
$\mathcal{M}$ & $\mathcal{U}(3,170)$ \\
$q$ & $\mathcal{U}(0.1,1)$ \\
$a_1,\,a_2$ & $\mathcal{U}(0,0.99)$ \\

\hline
\end{tabular}
\end{table}

We chose to study a population model that only depends on the masses, spins, and redshift so that we restrict $\theta \in \{m_1, m_2, a_1, a_2, z\}$ and we consider a selection function depending on the probability of detecting GW events as \cite{Selection}:
\begin{equation}
\frac{N_{\exp }(\Lambda)}{N(\Lambda)}=\int_{\rho \geq \rho_{\mathrm{th}}} p_{\mathrm{pop}}\left(m_{1 s}, m_{2 s}, z \mid \Lambda\right) \mathrm{d} m_1 \mathrm{~d} m_2 \mathrm{~d} z,
\end{equation}

This integral is evaluated via Monte Carlo integration over a set of $N_{\text{sim}}$ simulated injections drawn from an initial probability density function $\pi\left(m_{1i}, m_{2i}, z_i \mid \Lambda\right)$, of which $N_{\text{det}} $ are detected above the SNR threshold ($\rho \geq \rho_{\text{th}}$). Given this, the probability of detection is given by:

\begin{equation}
\begin{aligned} \frac{N_{\exp }(\Lambda)}{N(\Lambda)} & =\int_{\rho \geq \rho_{\text {th }}} p_{\text {pop }}\left(m_1, m_2, z \mid \Lambda\right) \mathrm{d} m_1 \mathrm{~d} m_2 \mathrm{~d} z, \\ & =\frac{1}{N_{\text {sim }}} \sum_{i=1}^{N_{\text {det }}} \frac{p\left(m_{1 i}, m_{2 i}, z_i \mid \Lambda\right)}{\pi\left(m_{1i}, m_{2 i}, z_i \mid \Lambda\right)}.\end{aligned} \end{equation}


 

\subsubsection{Results on population-level impact}
In this section, we explore the impact of waveform systematics on the BBH mass distribution hyperparameters using the Bayesian approach discussed in the last section. Before proceeding with the analysis on the simulated data with GW source parameters inferred using approximate waveform models, we validated the Bayesian code on mock samples where the dummy posteriors are considered Gaussian centred at the true value of the GW source parameters. We perform the hierarchical population analysis on these idealised Gaussian posteriors centred on the true injection parameters for the PL+G model. This test (see Appendix \ref{app:pipeline_validation} for details) recovered all population parameters to well within the 68\% credible interval, confirming that our code is robust.

\begin{figure*}[htbp]
    \centering    
    \begin{subfigure}{\linewidth}
        \centering
        \includegraphics[width=\linewidth]{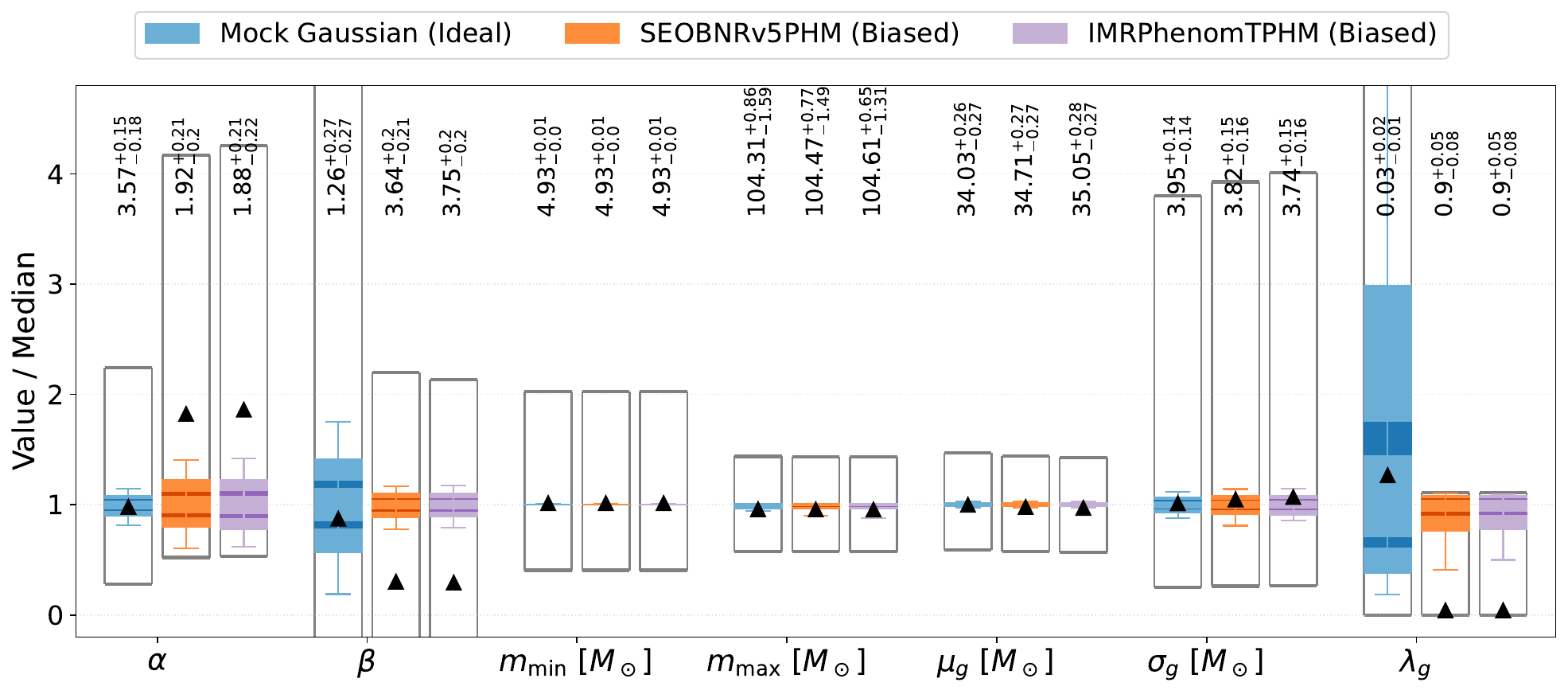}
        \label{fig:pop_plg_top}
    \end{subfigure}

    \caption{\justifying\textbf{Summary plot of hierarchical Bayesian inference results} The panel shows results for the Power Law+Gaussian population model where the blue box plots correspond to the idealised mock Gaussian posterior validation runs, while the orange and purple box plots show the same for the \texttt{SEOBNRv5PHM} and \texttt{IMRPhenomTPHM} recovery, respectively. The x-axis represents all the inferred hyperparameters, and all posterior samples are divided by their respective median values to allow comparison across parameters on a common $y$-axis. The darker-shaded boxes span $1\sigma$ of the data around the median, while the lighter-shaded boxes extend to $2\sigma$. Whiskers indicate the full range of the posterior samples. Black triangles mark the injected values divided by the corresponding median. The grey boxes around each hyperparameter refer to the prior bounds for each in units normalised by the individual posterior median. The text represents the median value of the parameter and the 68\% confidence interval around it (refer to Table \ref{tab:hyperparams_combined}).}
    \label{fig:summary_plot}
\end{figure*}

We then perform the analysis on the simulated GW source samples with posteriors inferred using an approximate waveform model. The results on the inferred hyperparameters are shown in Fig. \ref{fig:corner} and the summary plot showing the deviation in Figure \ref{fig:summary_plot}, along with the prior range.  Table \ref{tab:hyperparams_combined} summarises the recovered population hyperparameters for both mass models. We discuss below the key findings:

\begin{figure*}[htbp]
    \centering
    \includegraphics[width=1.05\linewidth]{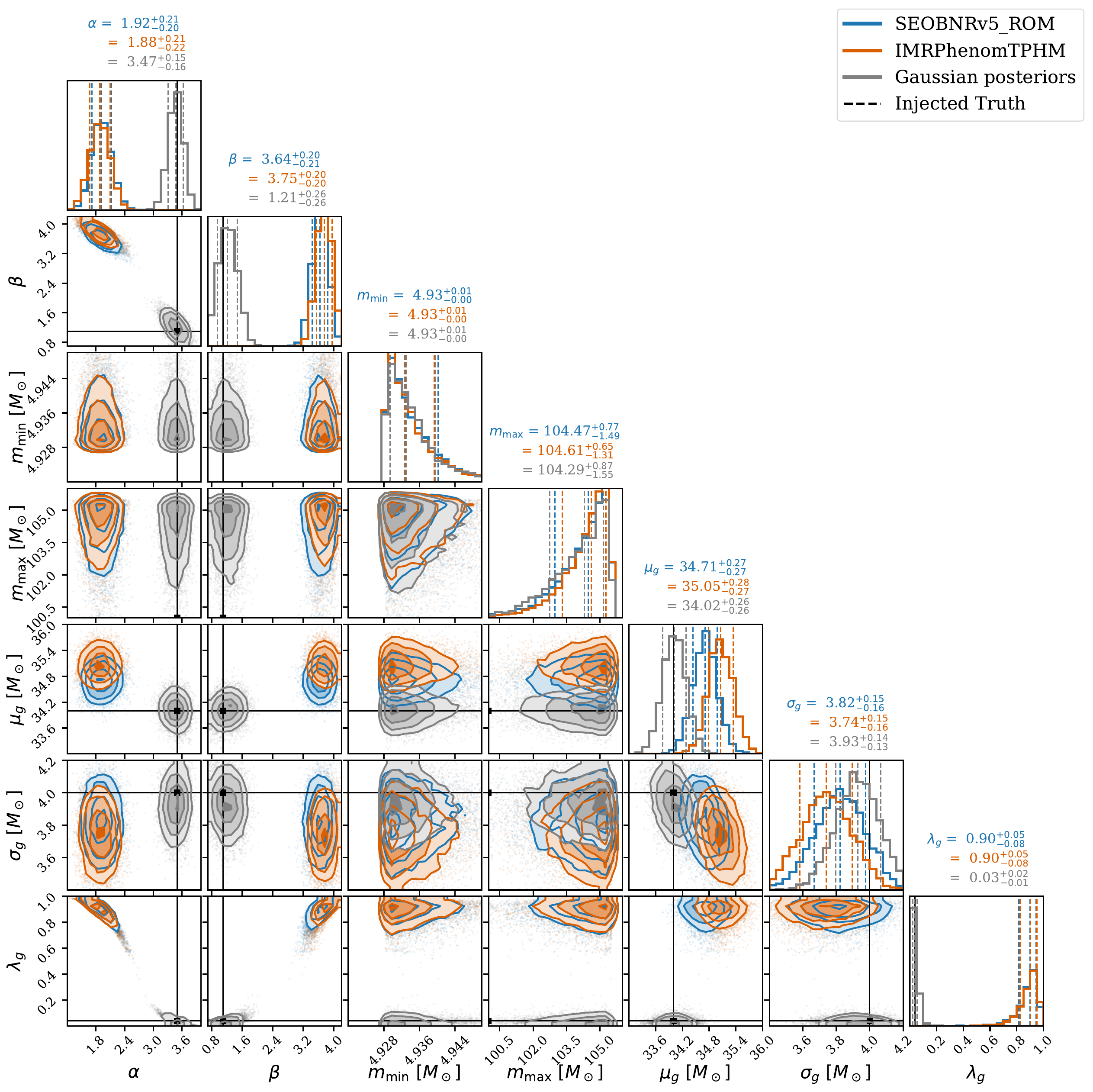} 
    \caption{\justifying \textbf{Hyperparameter posteriors from Hierarchical Bayesian Inference for the PL+G mass model.} One and two-dimensional marginalised posteriors for the population hyperparameters $\{\alpha,\,\sigma_g,\,\mu_g,\,\beta,\,m_{\min},\,m_{\max},\,\lambda_g\}$ are shown for recovery using \texttt{SEOBNRv5PHM} (blue), \texttt{IMRPhenomTPHM} (orange) and idealised Gaussian posteriors (grey). Dashed vertical lines in the 1D panels indicate the 68\% credible interval, and contours in the 2D panels show the 1$\sigma$, 2$\sigma$, and 3$\sigma$ regions. The truth lines are given in black.} 
    \label{fig:corner}
\end{figure*}


\begin{table}[t]
\centering
\caption{\justifying \textbf{Recovered population hyperparameters from hierarchical Bayesian inference.} Results are shown for \texttt{SEOBNRv5PHM} and \texttt{IMRPhenomTPHM} (in parentheses), and compared against the Gaussian posterior medians (Appendix~\ref{app:pipeline_validation}). The shift $\Sigma$, computed in column 4 is the difference between the recovered median (column 3) and the recovered Gaussian  median (column 3, Table V in Appendix B).} 
\label{tab:hyperparams_combined}
\renewcommand{\arraystretch}{1.45}
\begin{tabular}{l c c c}
\hline\hline
Parameter & True values 
 & Recovered & $\Sigma$ \\
\hline

$\alpha$ 
 & $3.50$
 & $1.92^{+0.21}_{-0.20} (1.88^{+0.21}_{-0.22})$  & $-1.55 (-1.59)$
  \\[3pt]

$\beta$ 
 & $1.10$
 & $3.64^{+0.19}_{-0.21} (3.74^{+0.15}_{-0.16})$ 
 & $2.43 (2.53)$
 \\[3pt]

$\mu_g\;[M_\odot]$ 
 & $34.0$
 & $34.71^{+0.27}_{-0.27} (35.05^{+0.28}_{-0.27})$ 
 &  $0.69(1.03)$
\\[3pt]

$\sigma_g\;[M_\odot]$ 
 & $4.0$
 & $3.82^{+0.15}_{-0.16} (3.74^{+0.15}_{-0.16})$ 
 & $-0.11 (-0.19)$
 \\[3pt]
 
$\lambda_g$ 
 & $0.038$
 & $0.89^{+0.047}_{-0.083} (0.90^{+0.047}_{-0.077})$ 
 & $0.856 (0.866)$
\\[3pt]

$m_{\min}\;[M_\odot]$ 
 & $5.0$ 
 & $4.93^{+0.0073}_{-0.0034} (4.93^{+0.0067}_{-0.0032})$ 
 & $0.0 (0.0)$ 
  \\[3pt]

$m_{\max}\;[M_\odot]$ 
 & $100.0$ 
 & $104.47^{+0.77}_{-1.49} (104.61^{+0.65}_{-1.31})$ 
 & $0.16(0.3)$
 \\[3pt]

\hline\hline
\end{tabular}
\end{table}

\begin{enumerate}
    \item \textbf{Gaussian peak location ($\mu_g$) and peak width ($\sigma_g$):} In the PL+G model, the peak of the Gaussian component, injected at 
    $\mu_g^{\mathrm{true}} = 34\,M_\odot$, is recovered at $\mu_g^{\mathrm{rec}} = 34.71\,M_\odot$, corresponding to a systematic shift of $\Delta\mu_g = 0.71\,M_\odot$. While the error bars in the measurement are small, it represents a $2.6 \sigma \text{ and } 3.8\sigma$ deviation for the SEOBNR and the IMRPhenom model respectively, from the injected value relative to the posterior uncertainty, indicating a statistically significant systematic shift. This shift is quantitatively consistent with the event-level bias expected at this mass scale. From the bias–mass scaling relation $|\Delta m_1| \propto m_1^\gamma$ with $\gamma \sim 0.81-0.92$, the expected systematic offset at $m \sim 34\,M_\odot$ is $\sim 0.5$–$1.5\,M_\odot$. The observed shift of $0.71\,M_\odot$ lies well within this range, demonstrating a direct propagation of event-level biases into population-level inference. Since the bias increases with mass, events near the injected Gaussian peak are, on average, shifted to higher recovered masses, collectively pulling the inferred peak location upward.
    The width of the Gaussian component in the PLG case is recovered as $\sigma_g^{\mathrm{rec}} = 3.82 \,M_\odot$, consistent with the injected value of $\sigma_g^{\mathrm{true}} = 4\,M_\odot$, well-within the $2\sigma$ level. Although the bias varies with mass, the relative variation across the Gaussian width is small, thus preserving it.

    \item \textbf{Gaussian fraction ($\lambda_g$):} The mixing fraction is recovered as $\lambda_g^{\mathrm{rec}} \sim 0.9$, an order-of-magnitude increase from the injected value of $0.038$. As seen in Fig. \ref{fig:bias_scaling_populations}, individual events are systematically recovered with higher masses, the shift being more in the high-mass regime. This results in an accumulation of posterior support in the region around $\sim 32$–$38\,M_\odot$, where the injected Gaussian component lies. The hierarchical model interprets this excess density not as a shifted population, but as evidence for a dominant Gaussian component, hence artificially inflating $\lambda_g$. A visual representation of this effect is given in Apendix \ref{app:pile_up}. 

    
    \item \textbf{Power-law slope $\alpha$ and $\beta$: } These parameters exhibit large, correlated biases, as we find $\alpha$ flattens from $3.5$ to $\sim 2$. This flattening is directly correlated with the inflated $\lambda_g$ (as is evident from Fig \ref{fig:corner}. The parameter $\beta$ steepens in both cases ($\approx 4$) versus the injected value of $1.1$, indicating that the model is driven towards extreme mass ratios. This is an artefact of the well-known $\mathcal{M}-q$ degeneracy: $\mathcal{M}$ is the best measured intrinsic parameter, meaning that a bias in the total mass is absorbed by a shift in $q$. This degeneracy becomes more pronounced with high-mass systems, where chirp mass estimation is also bad, thus showing up as a very steep $\beta$.
    
    \item \textbf{Boundary parameters $m_{\text{min}}$ and $m_{\text{max}}$:} The minimum mass is recovered accurately, while the maximum mass shows negligible shift with respect to the ideal case of a Gaussian posteriors, which itself show some deviation from the injected values. These boundary parameters are thus consequences of finite sampling and selection effects, rather than an artifact of waveform systematics. 
\end{enumerate}

Figure \ref{fig:corner} shows that while the Gaussian posterior analysis recovers the injected hyperparameters accurately, both waveform-based recoveries converge to similar but significantly biased parameters. We additionally find from the event-level parameter estimation, the inferred hyperparameters and the corresponding posteriors, both waveform families are biased similarly by comparable amounts irrespective of the discrepancies seen between them in Fig. \ref{fig:rec_vs_true_violin}. This agreement suggests that these biases are not because of a flaw in either waveform family but a structural limitation in how these approximate models capture the features for high mass events, relevant for the LVK frequency band.





\section{Discussion}%
\label{sec:Discussion}

This study establishes that waveform systematics introduce mass-dependent biases that grow as $|\Delta| \propto M^\gamma$, which directly propagate into population-level inference and distort spectral features in the recovered BBH mass distribution, which can now be predicted. 

In the PL+G model, we find that the Gaussian peak location $\mu_g$ is recovered with its location shifted by $\Delta \mu_g = 0.71\,M_\odot$, its width recovered accurately, however, its mixing fraction increased by more than an order of magnitude from $\lambda_g^{\mathrm{true}} = 0.038$ to $\lambda_g^{\mathrm{rec}} \approx 0.90$, a $\sim 13\sigma$ bias (Table \ref{tab:hyperparams_combined}). This indicates that although waveform systematics does not notably shift or broaden spectral features, it attributes nearly the entire population to the Gaussian component at the expense of the power law, and thus the correlated flattening of $\alpha$. $\beta$ also steepens due to the high bias in $m_1$, as $\mathcal{M}$ remains well-constrained, $q$ has to adjust for compensation in the opposite direction. $m_\text{max}$ is also overestimated at the $4\sigma$ level, trying to capture the tail of the overestimated posteriors of the individual events. Noteworthy is the precision with which the hierarchical model converges to these biased values, indicated by the narrow Gaussian posteriors, implying how the distortions mimic astrophysical measurements confidently. 

The practical implication is that current estimates of the PPISN feature and power-law slopes carry an unquantified systematic uncertainty due to possible systematics associated with approximate waveform models. Our study indicates that the trend in inferring a higher value of the hyperparameters from the injected ones in the simulation can actually significantly shadow the true value of the BBH population parameters due to waveform systematics. We plan to perform a study in the future on the GW-detected catalog to estimate this effect in more depth.  


Furthermore, we plan several extensions of this work in order to understand the impact of waveform systematics on the astrophysical and cosmological inferences. Here, we focused on the mass distribution, but the inferred spin distribution is likely subject to analogous waveform-systematics. The strong degeneracy between mass and spin \cite{Massspin, mass_spin,spinmass} in GW parameter estimation raises the question of the extent to which this observed correlation is physical, or an artefact of mass-dependent biases leaking into the spin posteriors through this degeneracy. A systematic injection-recovery study similar to the one presented here, but targeting the spin distribution, would address this question. A future study can involve testing whether non-parametric population models are more or less robust to these waveform systematics than the parameterised model used here. Nevertheless, we expect similar systematics to emerge because they have the flexibility to fit the data and thus are liable to reconstruct the shifted peaks and broadened features. 

Our results motivate several possible solutions to this problem. One of the computationally less costly solutions is to estimate the possible systematic uncertainty for different masses of the GW sources, and propagate that uncertainty in the population inference of the BBH mass distribution. We have shown the scaling of the uncertainty on mass in this work. Inclusion of this systematic uncertainty along with the statistical uncertainty on the inferred BBH mass distribution will make our findings robust. Also, for model-agnostic inference, introducing a discrete hyperparameter that can marginalize over the waveform uncertainty will be important. We plan to set up such formalism in an upcoming work. In the future, we will also explore the impact of waveform systematics on spin inference and how it correlates with the inferred masses of the GW sources. 

Other solutions include the development of more accurate approximate waveform models for high mass and unequal mass ratios. From our analyses themselves, we observe the consistency in recovery for the two waveform families we made use of. Thus, parameter estimation cannot be improved by just comparing across multiple approximate waveforms. Instead, it is pertinent to develop accurate waveform models and compare them with NR simulations. A recent study by \citet{moreEvent2} with NRSurrogate models found more than 20\% of events, recover noticeably different measurements of black hole properties like the masses and spins in comparison to other waveform models. 

However, compared to the approximate waveform models, surrogate models have a restricted regime of validity due to the unavailability of NR simulations over the entire parameter space. For example, the surrogate model in our analysis \texttt{NRSur7dq4} is valid only for $q \geq 1/6$ and total mass $\geq 60 M_\odot$ for a reference frequency of 20 Hz. Thus, such models calibrated to larger NR catalogs covering wider regions of the parameter space need to be developed. However, these methods can be computationally costly, given the expected large volume of GW source detections in the coming years. 

\section{Conclusion}%
\label{sec:Conclusion}
In this paper, we have quantified the systematic errors arising from the approximations in waveform modelling, establishing for the first time how event-level biases propagate to the population level and distort the inferred BBH mass distribution. Unlike statistical uncertainties, waveform systematics are parameter-dependent and thus shift the recovery in deterministic directions rather than introducing zero-mean scatter. Consequently, they do not average away with increasing catalog size. Our results highlight the need to incorporate waveform modelling uncertainty into the systematic error budget of population-level and cosmological analyses, especially as GW astronomy enters the precision era. Using an injection-recovery framework with \texttt{NRSur7dq4} as the fiducial true signal and \texttt{SEOBNRv5PHM} and \texttt{IMRPhenomXPHM} as recovery  templates, we arrive at the following conclusions:

\begin{enumerate}
    \item The systematic bias introduced by current waveform families follows a power law relation $|\Delta| \propto M^\gamma$ ($M \in \{m_i, \mathcal{M}\}$), due to the decreasing number of observable inspiral cycles at high mass, which shifts the signal power into the merger and ringdown phases where these models are less accurate. This scaling persists irrespective of controlled and realistic injection setups, confirming that it is an intrinsic property of the inter-model disagreement that affects real GW events.
    \item When propagated through hierarchical inference on a power law and Gaussian (PL+G) population, these event-level biases shift the Gaussian peak location by $0.71M_\odot$ and accurately recovers the Gaussian width. The most striking distortion is seen in the inflation of the Gaussian mixing fraction from the injected $\lambda_g = 0.038$ to $\sim 0.9$, the power-law slope flattens from injected $\alpha = 3.5$ to $\sim 2$ and mass-ratio index, to compensate, steepens from $\beta = 1.1$ to $\sim 4$. These biases are confidently measured by the hierarchical analysis, proving that waveform systematics can produce biases that look like well-constrained astrophysical measurements. Both waveform families employed show consistent deviations from injected values across both event and population-level stages of the analysis pipeline\footnote{We have also checked that our pipeline is able to reproduce the correct population parameters of BBHs when the event-level posteriors are not biased due to waveform systematics.}.
\end{enumerate}

The impact of waveform systematics can have far-reaching consequences beyond only the astrophysical population of BBHs. For science cases like inference of the expansion rate of the universe, GW lensing \cite{systematicsLensing, lensing}, and testing general relativity in the strong field regime, the impact of waveform systematics is essential to understand. For example, GW sources serve as 'standard sirens' providing a direct measurement of the luminosity distance $D_L$ that is independent of the cosmic ladder \cite{sirens, H0_gw}. When combined with redshift information, obtained either from an electromagnetic counterpart ('bright sirens'), from statistical cross-correlation with galaxy catalogs ('dark sirens') \cite{darkSirens,darkSirens2} or from the full redshifted mass distribution of the BBH mass spectrum ('spectral sirens')\cite{spectralSiren, spectralSiren2, SpectralSiren4, SpectralSiren3, SpectralSiren5}, this enables an independent measurement of the Hubble constant, offering a potential resolution to the long-standing Hubble tension problem. The spectral siren method, in particular, relies on exploiting the distinct features in the source-frame mass distribution, such as the location of peaks, cutoffs, or pair-instability supernova mass gap, and using their apparent shift in the detector frame to infer the redshift-distance relation. However, if waveform systematics introduce mass-dependent biases that distort these spectral features, as demonstrated by \citet{Hubble}, the resulting $H_0$ measurement can be impacted. Similarly, for GW lensing studies, waveform systematics lensing estimates \cite{lensing, systematicsLensing, Chakraborty:2025pxt} and for tests of general relativity in the strong-field regime, it is assumed that the GR waveform templates themselves are sufficiently accurate \cite{GRsystematics,lensingGR}. Systematic errors in the waveform can masquerade as apparent deviations from GR, leading to incorrect claims or, conversely, can create unnecessary degeneracies, hiding real deviations. As we enter an era where detector sensitivity will no longer be the limiting factor, understanding and quantifying waveform systematics becomes essential for the credibility of GW science. 

In summary, we would like to say that the success of GW astronomy depends strongly on the accuracy of the waveform modelling, not only for the detection of sources, but also for the accurate characterization of the underlying astrophysical population and deriving scientific outcomes useful for other branches of physics, such as cosmology and fundamental physics. Development of more accurate waveform models, which can be used in a source parameter inference code, will be essential for robust discovery of BBH mass distribution and discovering the underlying astrophysical population model and inferring fundamental physics from it.  

\section*{Acknowledgments}
The authors are thankful to Akash Maurya for carefully reviewing the manuscript as a part of the LSC publication and presentation policy and providing useful feedback. This work is a part of the ⟨Data$|$Theory⟩ Universe Lab, which is supported by the Department of Astronomy and Astrophysics at the Tata Institute of Fundamental Research (TIFR) and the Department of Atomic Energy, Government of India. We acknowledge the support of the Department of Atomic Energy, Government of India, under Project Identification No. RTI 4012. This research is supported by the Prime Minister Early Career Research Award, Anusandhan National Research Foundation, Government of India. We are also thankful for the computing resources provided by the ⟨Data$|$Theory⟩ Universe Lab. LIGO, funded by the U.S. National Science Foundation (NSF), and Virgo, supported by the French CNRS, Italian INFN, and Dutch Nikhef, along with contributions from Polish and Hungarian institutes. This collaborative effort is backed by the NSF’s LIGO Laboratory, a major facility fully funded by the National Science Foundation. We acknowledge the use of the following software packages in this work: \texttt{NumPy \cite{numpy}, SciPy \cite{scipy}, Pandas \cite{pandas}, Bilby \cite{bilby_paper} , GWSim \cite{GWSim}, LALSuite \cite{lalsuite}, Matplotlib \cite{matplotlib}, and AstroPy \cite{astropy}.}

\appendix

\section{Waveform models}\label{app:waveform}
Modelling the complete GW signal from a coalescing BBH, spanning inspiral, merger and ringdown, requires solving the full non-linear Einstein's field equations, a task that is very computationally expensive and time-consuming. To address this, several families of approximate waveform models have been developed. Here, we give a brief overview of three such classes: \texttt{IMRPhenom} phenomenological models, \texttt{SEOBNR} effective-one-body models and \texttt{NRSurrogate} numerical relativity surrogate models, highlighting their structural differences and the origin of waveform systematics studied in this paper.

\begin{itemize}
    \item \texttt{\textbf{IMRPhenom}}: The phenomenological models are modular, hybrid models that construct the waveform by connecting distinct piecewise regions, the inspiral part described by analytical post-Newtonian (PN) equations, and for the merger and ringdown, phenomenological fits calibrated to the full numerical relativity solutions are utilised. These independent regions are combined by enforcing continuity in the phase and amplitude.

Waveform systematics in this approach originate from several structural approximations. To reduce the dimensionality of the fits, aligned-spin models utilise effective single-spin parametrisations, which can degrade in accuracy for unequal-spin systems at high mass ratios. Further, because the phenomenological fits are built upon a finite set of NR simulations, the approach faces modelling errors in regions of parameter space that are outside or poorly sampled by the calibration bounds.

The latest generations of these models include \texttt{IMRPhenomXPHM} \cite{XPHM}, a frequency-domain model that incorporates both spin precession and higher-order multipoles through a "twisting-up" procedure applied to an aligned-spin baseline, 

To address these limitations, we've made use of its time-domain counterpart \texttt{IMRPhenomTPHM}, which models precession dynamics directly in the time-domain \cite{TPHM}.  It avoids the SPA and allows for the fully numerical evolution of the spin equations. However, it is worth noting that this is computationally more expensive than its frequency-domain counterpart, which is why they are more widely used for large-scale analyses.

\item \textbf{\texttt{SEOBNR}:} The spinning, effective one-body models map the complex two-body problem to the simpler problem of a single test mass moving in a deformed Kerr spacetime, where the deformation parameter is determined by the symmetric mass ratio of the binary. The structural framework of these models consists of three core building blocks: an EOB Hamiltonian describing the conservative dynamics that gives the equations of motion, the radiation-reaction force accounting for energy and angular momentum dissipation due to emitted GWs, and an inspiral-merger-ringdown waveform that combines analytical post-Newtonian inspiral information with merger-ringdown functional forms calibrated to numerical relativity (NR) data. 

The most widely used precessing, higher-multipole model in this family is \texttt{SEOBNRv5PHM} \cite{SEOB2}, and while this formalism is physically rigorous, waveform systematics still primarily arise from calibration limitations. 

\item \textbf{\texttt{NRSurrogate}: } Lastly, we have Numerical Relativity (NR) Surrogate models, which are not based on analytical approximations and are highly complex interpolations built directly from large catalogs of expensive NR simulations. They are the most accurate waveform families available, but are limited to specific regions of parameter space. Surrogate models take a set of pre-computed waveforms generated by an underlying model as input, in this case, the NR model, and interpolate in parameter space between these waveforms to produce waveforms for arbitrary parameter values \cite{NRSur}. NR simulations are direct numerical evolutions of fully-nonlinear Einstein's equations, which are expensive. Open NR waveform catalogs that enabled the development of such surrogate models include the NINJA-2 catalog \cite{ninja2}, the NRAR catalog \cite{NRAR}, and the SXS Public Catalog \cite{SXS}. In this work, we use the \texttt{NRSur7dq4} model \cite{NRSur_new} for injections, which includes all spin-weighted spherical harmonic modes up to $\ell=4$, as well as precession frame dynamics and spin evolution. 

In this work, we consider a signal generated using the NRSurrogate model to be the true GW signal and recover it using the other two classes of models described. 

\end{itemize}

\section{Hierarchical Inference Validation with Gaussian Posteriors}
\label{app:pipeline_validation}

To verify that the hierarchical inference pipeline itself is robust, we performed a controlled test in which Gaussian posteriors replace the individual-event posteriors centred on the true injected parameter values ($m_i \text{ and } a_i$), with $1\sigma$ widths. These Gaussian posteriors are then passed through the same hierarchical analysis code used for the real PE posteriors in Sec. \ref{sec:3}, with identical priors and sampler settings.

In table \ref{tab:gauss} we report the recovered population hyperparameters. All injected values are recovered within the 68\% credible intervals, confirming that the pipeline works as expected for well-behaved posteriors. The full posterior distribution for this validation run is shown in grey in Fig. \ref{fig:corner}. The one-dimensional marginals are unimodal and approximately Gaussian, while the two-dimensional contours show the recovered hyperparameters falling well within the 68\% credible region. 

\begin{table}[t]
\centering
\caption{\justifying \textbf{Gaussian-posterior recovery of PL+G hyperparameters.} The true injected values are compared against the median and 68\% credible interval recovered by the hierarchical pipeline when idealised Gaussian posteriors are used in place of the real PE samples. The bias $\Delta \equiv \Lambda_{\mathrm{rec}} - \Lambda_{\mathrm{true}}$ is sub-percent for all shape parameters, confirming that the pipeline is robust.}
\label{tab:gauss}
\renewcommand{\arraystretch}{1.35}
\begin{tabular}{l c c r}
\hline\hline
Parameter & True & Recovered & \multicolumn{1}{c}{$\Delta$} \\
\hline

$\alpha$ 
 & $3.50$ 
 & $3.47^{+0.15}_{-0.16}$ 
 & $-0.03$ \\[2pt]

$\beta$ 
 & $1.10$ 
 & $1.21^{+0.26}_{-0.26}$ 
 & $+0.11$ \\[2pt]

$\mu_g\;[M_\odot]$ 
 & $34.00$ 
 & $34.02^{+0.26}_{-0.26}$ 
 & $+0.02$ \\[2pt]

$\sigma_g\;[M_\odot]$ 
 & $4.00$ 
 & $3.93^{+0.14}_{-0.13}$ 
 & $-0.07$ \\[2pt]

$\lambda_g$ 
 & $0.038$ 
 & $0.034^{+0.022}_{-0.013}$ 
 & $-0.004$ \\[2pt]

$m_{\min}\;[M_\odot]$ 
 & $5.00$ 
 & $4.93^{+0.01}_{-0.00}$ 
 & $-0.07$ \\[2pt]

$m_{\max}\;[M_\odot]$ 
 & $100.0$ 
 & $104.29^{+0.87}_{-1.55}$ 
 & $+4.29$ \\[2pt]

\hline\hline
\end{tabular}
\end{table}


This check establishes two points: 
\begin{enumerate}
    \item When provided with well-behaved posteriors centred on the truth, the hierarchical analysis recovers all population hyperparameters to within statistical uncertainties. 
    \item Most importantly, the biases in the hyperparameters reported in the main analysis are physical and arise entirely from the systematic mismatch between injection and recovery waveforms.
\end{enumerate}

\section{Visualising the Artificial Mass Pile-Up}
\label{app:pile_up}

\begin{figure}[htbp]
    \centering
    \includegraphics[width=\linewidth]{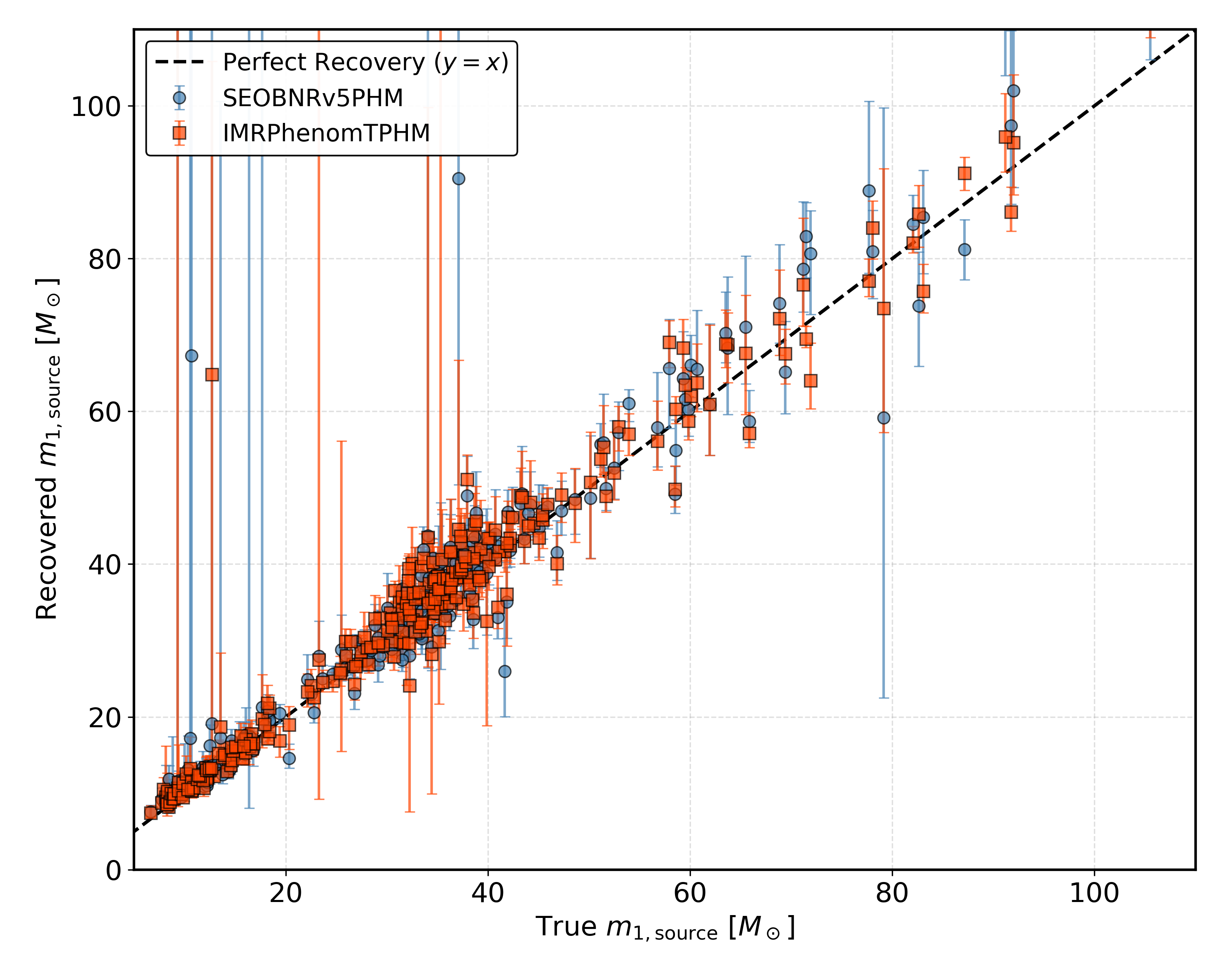}
    \caption{\justifying Events from the injected PL+G catalog are plotted with their injected source-frame primary mass versus their recovered primary mass using \texttt{SEOBNRv5PHM} (blue circles) and \texttt{IMRPhenomTPHM} (orange squares). The dashed $y=x$ line shows perfect recovery.}
    \label{fig:source_rec_scatter}
\end{figure}

To visually demonstrate the origin of the overestimation in the Gaussian mixing fraction ($\lambda_g$), we plot the median recovered source-frame primary mass against the injected source-frame mass for the entire PLG mock catalog (Fig. \ref{fig:source_rec_scatter}). 

The masses below $20M_\odot$ are recovered well and closely track the diagonal line $y=x$. For true masses between $m_{1,\mathrm{source}} \sim 25-40 \,M_\odot$, we start observing a distinct directional bias which was also explicit in the P-P plot in \ref{fig:pp_plot}. The systematic upward shift forces a large number of events to compress into a dense, horizontal band centred around recovered $m_{1,\mathrm{source}} \approx 34 M_\odot$. The detector sensitivity also peaks in the $30–40\,M_\odot$ range in source frame (after accounting for the selection function), so more events pass the SNR threshold and enter the analysis in this region. The combination of these two effects is that the hierarchical parameter estimation absorbs this artificially dense band of over-estimated masses by inflating the Gaussian peak component, providing a direct physical justification for why the mixing fraction $\lambda_g$ jumps from its true value of $0.038$ to  $\sim 0.9$.

\bibliographystyle{apsrev4-1}
\bibliography{main_paper.bib}

\end{document}